\documentclass[aps,prd,onecolumn,preprintnumbers]{revtex4}

\usepackage{graphicx}
\usepackage{dcolumn}
\usepackage{bm}
\usepackage{amsmath}
\usepackage{amssymb}
\usepackage{amsfonts}
\usepackage{float}
\usepackage{hyperref}[11in]
\usepackage{dsfont}
\usepackage{slashed}
\usepackage{booktabs}
\usepackage{multirow}
\usepackage{subfigure}
\usepackage[sort&compress]{natbib}
\usepackage{xcolor}
\usepackage{color}
\usepackage{colordvi}
\usepackage{wasysym}
\usepackage{array,multirow}
\usepackage{booktabs}
\usepackage{bbold}

\usepackage{graphicx}
\usepackage{xfrac}

\newcommand{\be}{\begin{equation}}
\newcommand{\ee}{\end{equation}}
\newcommand{\beq}{\begin{eqnarray}}
\newcommand{\eeq}{\end{eqnarray}}

\newcommand{\tr}{\mathrm{Tr}}

\begin{document}
\title{Quark flavor decomposition of the nucleon axial form factors}

\author{
  C.~Alexandrou$^{1,2}$,
  S.~Bacchio$^{2}$,
  M.~Constantinou$^{3}$,
  K.~Hadjiyiannakou$^{1,2}$,
  K.~Jansen$^{4}$,
  G.~Koutsou$^{2}$
}
\affiliation{
  $^1$Department of Physics, University of Cyprus, P.O. Box 20537, 1678 Nicosia, Cyprus\\
  $^2$Computation-based Science and Technology Research Center, The Cyprus Institute, 20 Kavafi Str., Nicosia 2121, Cyprus \\
  $^3$Department of Physics, Temple University, 1925 N. 12th Street, Philadelphia, PA 19122-1801,
USA\\
  $^4$NIC, DESY, Platanenallee 6, D-15738 Zeuthen, Germany\\
}

\begin{abstract}
  \centerline{\includegraphics[width=0.15\linewidth]{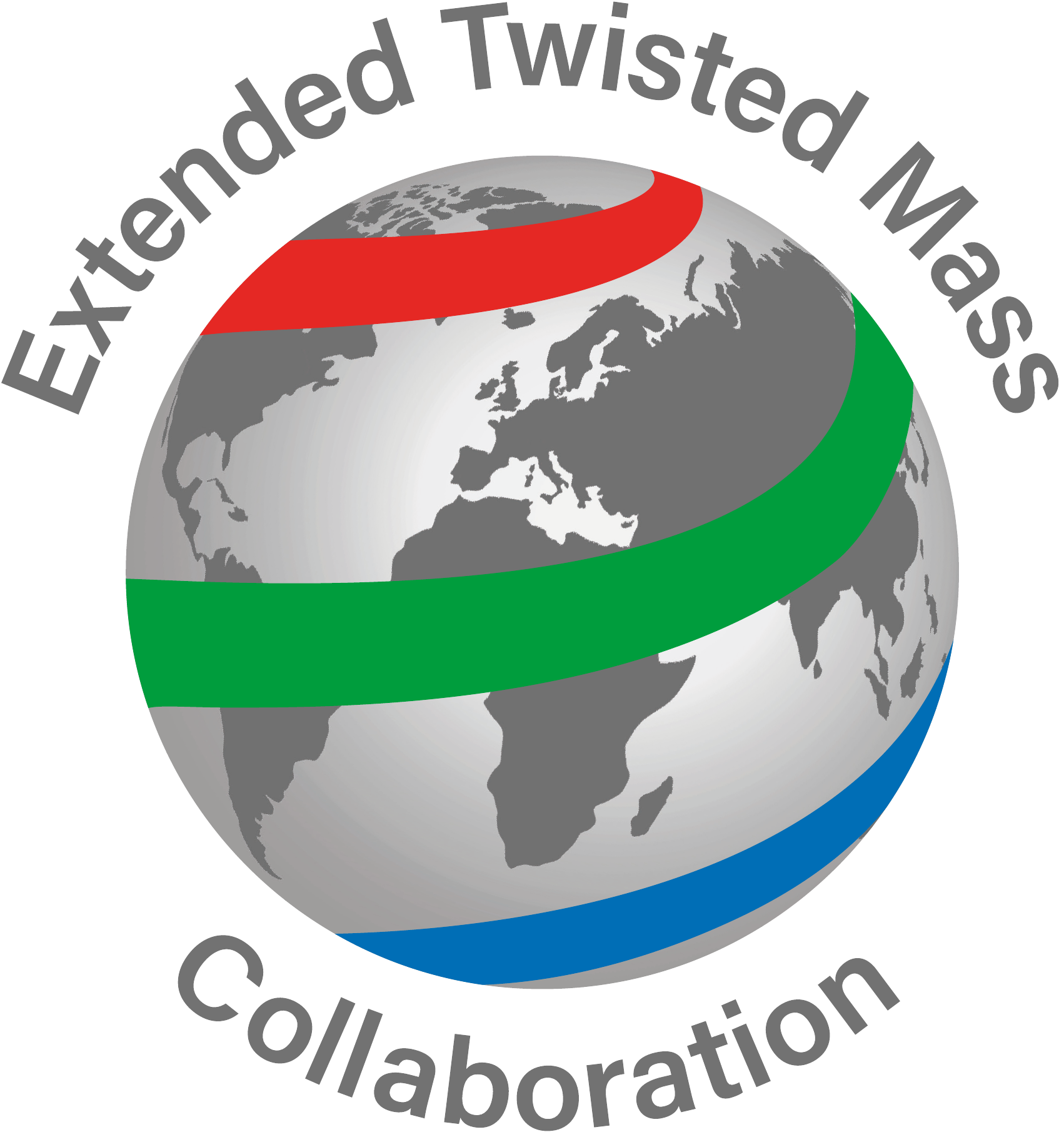}}
  \vspace*{0.3cm}
 We present results on the isoscalar form factors including the disconnected contributions, as well as on the strange and charm quark form factors. Using previous results on the isovector form factors, we determine   the flavor decomposition of the nucleon axial form factors. These are computed using an ensemble of $N_f=2+1+1$ twisted mass fermions simulated with physical values of quark masses. We investigate the SU(3) flavor symmetry and show that there is up to 10\% breaking for the axial  and up to 50\% for the induced pseudoscalar form factors. By fitting the $Q^2$-dependence, we determined the corresponding  root mean square radii. The pseudoscalar coupling of the $\eta$ meson and the nucleon is found to be $g_{\eta NN}=3.7(1.0)(0.7)$, and the  Goldberger-Treiman discrepancy for the octet combination about 50\%.
\end{abstract}

\maketitle
\bibliographystyle{apsrev}

\section{Introduction}
Axial form factors play a key role in the interactions of nucleons with the W and Z bosons, the carriers of the weak force. They also provide insights into the structure of the nucleon that in turn can affect our ability to compute cross sections that may aid us into revealing new physics.   Neutron beta decay and other charged current weak interaction  processes like $\nu_\mu+n\rightarrow p+\mu^-$ are  sensitive to the isovector axial form factor $G_A^{u-d}(Q^2)$. Neutrino elastic scattering on protons is sensitive to the strange axial form factor of the proton $G_A^s(Q^2)$, which for $Q^2=0$ determines the strange quark contribution to the proton spin $\Delta s$. The role of strange quarks is also important for calculating the cross sections for a class of popular cold dark matter candidates ~\cite{Papavassiliou:2009zz}.  A variety of experiments ranging from nuclear recoil direct-detection experiments to collider indirect-experiments are searching for dark matter candidates that use as input either spin-dependent or spin-independent nucleon cross sections. A first  measurement of parity-violating asymmetries in forward elastic electron-proton scattering by HAPPEx~\cite{Aniol:2004hp} combined with  data from neutrino and antineutrino-proton  elastic scattering cross sections from Brookhaven E734~\cite{Ahrens:1986xe} determined simultaneously the
strange vector and axial form factors of the proton at non-zero momentum transfer square $Q^2$~\cite{Pate:2003rk}.  Additional parity-violating data from the G0 experiments~\cite{Armstrong:2005hs,Androic:2009aa} improved the determinations of the strange axial form factors~\cite{Pate:2008va}.   The  MicroBooNE neutrino detector at Fermilab aims to extract the strange axial form factor of the nucleon in the range of momentum transfers  of 1~GeV$^2$ to as low as 0.08~GeV$^2$~\cite{Miceli:2014hva,Kim:2019mdm}. Combining neutrino-proton neutral and charged current scattering cross section measurements with available polarized electron-proton/deuterium cross section data is expected to reduce the experimental uncertainty and allow for the extraction of $\Delta s$ with an order of magnitude better accuracy complementing  polarized deep inelastic scattering experiments. The axial form factors are the main source of error in the description of neutrino-nucleon interactions.  Therefore, a calculation of these form factors within lattice QCD will provide valuable input in experiments such as DUNE~\cite{Abi:2020kei,Abi:2020qib}  and Hyper-K~\cite{Abe:2020hnh,Abe:2020sbr}.

 Lattice QCD  provides the {\it ab initio} non-perturbative framework  for computing the nucleon axial form factors using directly the QCD Lagrangian. While there are a number of lattice QCD studies of the isovector axial form factors with recent results given in Refs.~\cite{Alexandrou:2020okk,Jang:2019vkm, Rajan:2017lxk,Capitani:2017qpc,Alexandrou:2017hac,Green:2017keo}, only a few studies are done for other flavor combinations~\cite{Green:2017keo,Alexandrou:2017hac,Djukanovic:2019gvi}. 
 The reason for this is that the isovector flavor combination is free of  quark disconnected contributions. In Ref.~\cite{Alexandrou:2020okk} we presented our results for the isovector axial form factors, while also investigating finite volume effects.
 This work focuses on the study of the isoscalar, octet and singlet flavor combination by computing all disconnected contributions, allowing us to perform a flavor decomposition.
The computation is performed using one ensemble of  $N_f=2+1+1$ dynamical quarks with the up, down, strange and charm masses tuned to their physical values, referred to as physical point.

 The remainder of this paper is organized as follows: In Section~\ref{sec:A-V_ME} we discuss the PCAC  relation and the parameterization of the $Q^2$ dependence of the form factors. In Section~\ref{sec:LatM} we explain in detail the lattice
 methodology  to extract the nucleon axial and induced pseudoscalar form factors. In Section ~\ref{sec:Renormalization} we discuss the renormalization and in Section~\ref{sec:isos} we show results for  the isoscalar 
  combination, where both connected and quark disconnected contributions are presented. The  strange and charm form factors are presented in Section~\ref{sec:SC_FFs}, in Section~\ref{sec:fs_octet_comps}  the flavor singlet and octet combinations are discussed and in Section~\ref{sec:FL_decomp} we provide the results for  the form factors for each quark  flavor. Final results are quoted in Section~\ref{Sec:FinalRes}, comparisons with previous studies are carried out in Section~\ref{sec:Comp} and in Section~\ref{sec:Summary} we conclude. 
 
\section{Matrix elements, form factors and $Q^2$-dependence}\label{sec:A-V_ME}
In a previous paper~\cite{Alexandrou:2020okk}, we presented results on the isovector axial form factors $G_A^{u-d}(Q^2)$ and $G_P^{u-d}(Q^2)$, as well as, the pseudoscalar $G_5^{u-d}(Q^2)$. We refer the reader to that paper for details on the computation of the isovector combination. In this paper, we will describe the flavor combinations where disconnected contributions are involved, such as the isoscalar combination
\begin{equation}
    A^{u+d} = \bar{u} \gamma_\mu \gamma_5 u + \bar{d} \gamma_\mu \gamma_5 d.
    \label{Eq:AVcurrentIsos}
\end{equation}
Combining the isovector and isoscalar matrix elements one can extract the axial form factors for the up and down quarks. We will also compute the strange and charm form factors and construct SU(3) flavor combinations.
Considering the $u$, $d$ and $s$ flavor triplet we form the flavor singlet combination given by 
\begin{equation}
    A_\mu^0 \equiv A_\mu^{u+d+s} = \bar{u} \gamma_\mu \gamma_5 u + \bar{d} \gamma_\mu \gamma_5 d + \bar{s} \gamma_\mu \gamma_5 s.
    \label{Eq:AVcurrentSinglet}
\end{equation}
and the flavor octet  given by
\begin{equation}
    A_\mu^8 \equiv A_\mu^{u+d-2s} = \bar{u} \gamma_\mu \gamma_5 u + \bar{d} \gamma_\mu \gamma_5 d - 2\bar{s} \gamma_\mu \gamma_5 s,
    \label{Eq:AVcurrentOctet}
\end{equation}
In the SU(3) flavor symmetric limit, the matrix elements of $A_\mu^8$ will only have connected contributions. The axial Ward-Takahashi 
identity that leads to the  partial conservation of the axial-vector current (PCAC) is  
\begin{equation}
  \partial^\mu A^8_\mu= 2 i m_q P^8,
  \label{Eq:PCAC}
\end{equation}
where $P^8$ is the octet pseudoscalar density. The octet combination of the induced pseudoscalar form factor $G_P^{u+d-2s}$ is related to the pseudoscalar coupling between the $\eta-$meson and the nucleon.

The isosinglet flavor combination, on the other hand, has an anomalous term~\cite{Adler:1969gk} and it satisfies a modified relation,
\begin{equation}
    \partial^\mu A_\mu^0 = 6 {\cal Q} + 2im_q P^0
    \label{Eq:DivA0}
\end{equation}
where  $P^0= \bar{u} \gamma_5 u + \bar{d}  \gamma_5 d + \bar{s}  \gamma_5 s$ is the isosinglet  pseudoscalar  current, ${\cal Q}(x)$ is the topological density ${\cal Q}(x) = \frac{1}{32 \pi^2} \epsilon_{\mu\nu\rho\sigma} {\rm Tr} [F_{\mu\nu}(x) F_{\rho\sigma}(x)]$ and $F_{\mu\nu}$ is the field strength tensor of QCD. The anomalous gluonic term is induced by the axial anomaly. Since gluons couple equally to each quark flavor, the anomalous term vanishes only for non-singlet combinations as in Eqs.~(\ref{Eq:PCAC}).
The anomaly term has the consequence that the axial-vector flavor singlet current is not conserved even for massless quarks. 

The nucleon matrix element of the axial operators in Eqs.~\eqref{Eq:AVcurrentIsos}, ~\eqref{Eq:AVcurrentOctet} and ~\eqref{Eq:AVcurrentSinglet}  can be written in terms of the axial, $G_A(Q^2)$, and induced pseudoscalar, $G_P(Q^2)$, form factors as
\begin{equation}
    \langle N(p',s') \vert A_\mu\vert N(p,s) \rangle = \bar{u}_N(p',s')  
     \bigg[\gamma_\mu G_A(Q^2) - \frac{Q_\mu}{2 m_N} G_P(Q^2)\bigg] \gamma_5 u_N(p,s),
    \label{Eq:DecompA}
\end{equation}
where $u_N$ is the nucleon spinor with initial (final) momentum $p(p')$ and spin $s(s')$, $q=p'-p$ the momentum transfer and $q^2=-Q^2$. The expression is given in Euclidean space. Note that we have suppressed the index denoting the flavor combination for simplicity.

Calculations on the lattice allow us to compute the form factors only at given discrete values of  $Q^2$. In order to investigate their full $Q^2$ dependence we use two fit forms, the well known dipole Ansatz, and the model
independent z-expansion~\cite{Hill:2010yb,Bhattacharya:2011ah}.
In the case of the dipole Ansatz we have that
\begin{equation}
    G(Q^2) = \frac{G(0)}{(1+\frac{Q^2}{m^2})^2}.
    \label{Eq:dipole}
\end{equation}
In the case of the axial form factor, $G(0)$ gives the axial charge and $m$ the axial mass for the flavor combinations under investigation. The radius  is extracted 
from the slope  in the limit $Q^2\rightarrow 0$, namely
\begin{equation}
    \langle r^2 \rangle = - \frac{6}{ G(0)} \frac{d G(Q^2)}{dQ^2} \bigg \vert_{Q^2 \rightarrow 0}.
    \label{Eq:radius}
\end{equation}
Combining Eq.~(\ref{Eq:dipole}) and Eq.~(\ref{Eq:radius}) one can show that the radius
is connected to the dipole mass as
\begin{equation}
    \langle r^2 \rangle = \frac{12}{m^2}.
    \label{Eq:rToM}
\end{equation}
Customarily, one characterizes the size of a hadron probed by a given current  by the  root mean square radius (r.m.s) defined as $\sqrt{\langle r^2 \rangle}$. 

In the case of the z-expansion, the form factor is expanded in a series as,
\begin{equation}
    G(Q^2) = \sum_{k=0}^{k_{\rm max}} a_k\; z^k(Q^2),
    \label{Eq:zExp}
\end{equation}
where 
\begin{equation}
    z(Q^2) = \frac{\sqrt{t_{\rm cut} + Q^2} - \sqrt{t_{\rm cut}} }{ \sqrt{t_{\rm cut} + Q^2} + \sqrt{t_{\rm cut}} }
\end{equation}
imposing analyticity constrains, with $t_{\rm cut}$ the particle production threshold. We use the three-pion cut, $t_{\rm cut}=9 m_\pi^2$~\cite{Bhattacharya:2011ah} for all flavor combinations, although  apart from the isovector case,  the  cut-off might be higher due to heavier decay modes.
The coefficients $a_k$ appearing in Eq.~\eqref{Eq:zExp} should have an upper bound, so that the series converges at some value of $k$. Larger values of $a_k$ that could appear for $k>1$ can lead to instabilities. Therefore, we employ Gaussian priors, which are centered around zero with a chosen standard deviation $w\max(|a_0|,|a_1|)$~\cite{Green:2017keo}, and with $w$ controlling the width of the prior. The value of the form factor at zero momentum is $a_0$ and the radius is
\begin{equation}
    \langle r^2 \rangle = - \frac{3 a_1}{2 a_0 t_{\rm cut}}.
\end{equation}
The coefficients 
$a_0$ and $a_1$ are anticipated to have opposite signs in order to lead to positive values of the radii.
If we compare the above equation with the one extracted for the dipole fit of Eq.~(\ref{Eq:rToM})  we can define the corresponding mass determined from the z-expansion to be
  \be
  m=\sqrt{-\frac{8a_0t_{\rm cut}}{a_1}}.
  \ee
This relation will allow us to compare the radius extracted from the dipole and the z-expansion.

\section{Lattice methodology}\label{sec:LatM}
This section explains the methodology we use within lattice QCD in order to compute correlation functions and ensure  ground state dominance. It also provides details on the gauge ensemble used in the analysis.

\subsection{Correlation functions} \label{ssec:CorrFuncs}
For the computation of the correlation functions we use the standard nucleon interpolating field 
\begin{equation}
    {\cal J}_N(t,\vec{x}) = \epsilon^{abc} u^a(x) \left[ u^{b T}(x) \mathcal{C} \gamma_5 d^{c} (x) \right],
    \label{Eq:IntF}
\end{equation}
where $\mathcal{C}=\gamma_0 \gamma_2$ is the charge conjugation matrix and $u(x)$, $d(x)$ the up and down quark fields. The two-point function in momentum space is then expressed as
\begin{equation}
C(\Gamma_0,\vec{p};t_s,t_0) =  \sum_{\vec{x}_s} \hspace{-0.1cm} e^{{-}i (\vec{x}_s{-}\vec{x}_0) \cdot \vec{p}} \;\tr \left[ \Gamma_0 {\langle}{\cal J}_N(t_s,\vec{x}_s) \bar{\cal J}_N(t_0,\vec{x}_0) {\rangle} \right],
\label{Eq:2pf}
\end{equation}
where with $x_0$ we denote  the source and $x_s$ the sink positions on the lattice where states with the quantum numbers of the
nucleon are created and destroyed, respectively. 
$\Gamma_0$ is the unpolarized positive parity projector $\Gamma_0 = \frac{1}{2}(1+\gamma_0)$. 

For the construction of the three-point correlation function the axial-vector current is inserted at a time slice, $t_{\rm ins}$, between the time of the  creation and annihilation of states. The three-point function is given by 
\begin{equation}
C_{\mu}(\Gamma_\rho,\vec{q},\vec{p}\,';t_s,t_{\rm ins},t_0) =
 \hspace{-0.1cm} {\sum_{\vec{x}_{\rm ins},\vec{x}_s}} \hspace{-0.1cm} e^{i (\vec{x}_{\rm ins} {-} \vec{x}_0)  \cdot\vec{q}}  e^{-i(\vec{x}_s {-} \vec{x}_0)\cdot \vec{p}\,'} \; \tr \left[ \Gamma_\rho \langle {\cal J}_N(t_s,\vec{x}_s) A_{\mu}(t_{\rm ins},\vec{x}_{\rm ins}) \bar{\cal J}_N(t_0,\vec{x}_0) \rangle \right],
  \label{Eq:3pf}
\end{equation}
with $\Gamma_\rho$ is the polarized projector, $\Gamma_\rho = i \Gamma_0 \gamma_5 \gamma_\rho$ . 

\subsection{Ground state dominance}\label{ssec:ExcStates}
The interpolating field of Eq.~(\ref{Eq:IntF}) creates the nucleon ground state but also excited states. We apply Gaussian smearing~\cite{Alexandrou:1992ti,Gusken:1989qx} to the quark fields entering the interpolating field in order to increase the overlap with the ground state. See Ref.~\cite{Alexandrou:2020okk} for more details about our smearing procedure.
To isolate the matrix element of interest  we construct a ratio
of three- to a combination of two-point functions~\cite{Alexandrou:2013joa,Alexandrou:2011db,Alexandrou:2006ru,Hagler:2003jd}
\begin{equation}
R_{\mu}(\Gamma_{\rho},\vec{p}\,',\vec{p};t_s,t_{\rm ins}) = \frac{C_{\mu}(\Gamma_\rho,\vec{p}\,',\vec{p};t_s,t_{\rm ins}\
)}{C(\Gamma_0,\vec{p}\,';t_s)} \times
\sqrt{\frac{C(\Gamma_0,\vec{p};t_s-t_{\rm ins}) C(\Gamma_0,\vec{p}\,';t_{\rm ins}) C(\Gamma_0,\vec{p}\,';t_s)}{C\
(\Gamma_0,\vec{p}\,';t_s-t_{\rm ins}) C(\Gamma_0,\vec{p};t_{\rm ins}) C(\Gamma_0,\vec{p};t_s)}}.
\label{Eq:ratio}
\end{equation}
Overlap terms and time decaying exponentials cancel in the ratio. In Eq.~\eqref{Eq:ratio} and from now on, we consider that  $t_s$ and $t_{\rm ins}$ are expressed relative to the source $t_0$ i.e. $t_s-t_0\rightarrow t_s$ and $t_{\rm ins}-t_0\rightarrow t_{\rm ins}$. The ratio of Eq.~(\ref{Eq:ratio}) leads to the nucleon matrix element in the large time $t_s$ and $t_{\rm ins}$ limits, that is
\begin{equation}
  R_{\mu}(\Gamma_\rho;\vec{p}\,',\vec{p};t_s;t_{\rm ins})\xrightarrow[t_{\rm ins} \Delta E\gg 1]{(t_s-t_{\rm ins}) \Delta E\gg 1}\Pi_{\mu}(\Gamma_\rho;\vec{p}\,',\vec{p})\,,
\end{equation} 
where $\Delta E$ is the energy gap between the first excited state and the nucleon state. The rate of convergence to the nucleon state depends, besides the smearing procedure, also on the type of the insertion operator. In order to ensure ground state dominance, we employ three methods, namely a one state fit (plateau method), a two-state fit and the summation method. For a more detailed description about those three methods we refer the reader to  Ref~\cite{Alexandrou:2020okk}.

In this analysis we consider the same energy spectrum decomposition in both the two- and three-point functions. We determine the  first excited state energy $E_1(\vec {p})$ for each value of $\vec{p}$   by fitting the  two-point function  and use it when fitting the ratio of Eq.~(\ref{Eq:ratio}). We also fit the zero momentum two-point function to extract  the nucleon mass and then use the continuum dispersion relation to determine the lowest state energy $E_0(\vec{p}) = \sqrt{m_N^2 + \vec{p}^2}$  for a given value of momentum. As shown in Ref.~\cite{Alexandrou:2020okk}, the continuum dispersion relation is satisfied  for all the momenta considered in this work. 

From the nucleon  matrix element one can determine the axial $G_A(Q^2)$ and induced pseudoscalar $G_P(Q^2)$ form factors using the decomposition of Eq.~\eqref{Eq:DecompA}. Since there are several combinations of insertion, projector indices and momenta there is an over-constrained system of equations which determines  the form factors. Details are given in Appendix B of Ref.~\cite{Alexandrou:2020okk}.

\subsection{Ensemble of gauge configurations}\label{ssec:EnsDetails}
This work is based on the analysis of  an $N_f=2+1+1$ twisted mass clover-improved fermion ensemble,  referred to as  cB211.072.64 (see Table~\ref{table:sim}).
In Ref.~\cite{Alexandrou:2020okk}, where we studied the isovector form factors,  we also analyzed two $N_f=2$ ensembles with the same light quark action, namely the cA2.09.48 and cA2.09.64 ensembles  that have the same lattice spacing but different volumes (see Table ~\ref{table:sim}). Results on the axial form factors for the cA2.09.48 ensemble were first presented in Ref.~\cite{Alexandrou:2017hac}. For all the ensembles the lattice spacing is determined using the nucleon mass. More details are given in Refs.~\cite{Alexandrou:2018egz,Alexandrou:2018sjm,Alexandrou:2017xwd,Alexandrou:2019ali}. Finite volume effects have not been detected within the  statistical precision obtained for the isovector quantities and therefore we do not consider them here.

These gauge  configurations are produced by the Extended Twisted Mass Collaboration (ETMC) using the twisted mass fermion formulation~\cite{Frezzotti:2000nk,Frezzotti:2003ni} with a clover term~\cite{Sheikholeslami:1985ij} and the Iwasaki~\cite{Iwasaki:1985we} improved gauge action.  Since the simulation is done at maximal twist, we have automatic ${\cal O}(a)$ improvement for the physical observables studied here.

   \begin{center}
 \begin{table}[ht!]
  \caption{The parameters of the simulation for the $N_f=2+1+1$ cB211.072.64 ensemble~\cite{Alexandrou:2018egz} but also the two $N_f=2$ ensembles cA2.09.48~\cite{Abdel-Rehim:2015pwa} and cA2.09.64. $c_{SW}$ is the value of the clover
    coefficient and $\beta=6/g$ where $g$ is the bare coupling constant.  $N_f$ is the number of dynamical quark flavors, the lattice spacing is $a$ and
    the lattice volume is $V$. 
    $m_\pi$ is the pion mass and $m_N$ the nucleon mass. $L$  the spatial lattice length in physical units.}
  \label{table:sim}
       \begin{tabular}{l r@{.}l r@{.}l l  r@{$\times$}l c r@{.}l ccc r@{.}l r}
    \hline\hline
    Ensemble &\multicolumn{2}{c}{$c_{\rm SW}$} & \multicolumn{2}{c}{$\beta$} & \multicolumn{1}{c}{$N_f$} & \multicolumn{2}{c}{V} & $m_\pi L$ & \multicolumn{2}{c}{$a$ [fm]} & $m_N/m_\pi$ & $a m_\pi$ & $a m_N$ & \multicolumn{2}{c}{$m_\pi$ [GeV]} &$L$ [fm] \\
    \hline
    cB211.072.64 & 1&69    & 1&778 & 2+1+1 & $64^3$&$128$ & 3.62  & 0&0801(4)    & 6.74(3)    & 0.05658(6)   & 0.3813(19)      & 0&1393(7)          &5.12(3) \\
    cA2.09.64 & 1&57551 & 2&1   & 2        & $64^3$&$128$ & 3.97  & 0&0938(3)(1) & 7.14(4)    & 0.06193(7)   & 0.4421(25)      & 0&1303(4)(2)       &6.00(2) \\
    cA2.09.48 & 1&57551 & 2&1   & 2        & $48^3$&$96$  & 2.98  & 0&0938(3)(1) & 7.15(2)    & 0.06208(2)   & 0.4436(11)      & 0&1306(4)(2)       &4.50(1) \\
    \hline\hline
  \end{tabular}

 \end{table}
  \end{center}

\subsection{Disconnected three-point functions and statistics}\label{ssec:ConnDiscStats}
The three-point function defined in Eq.~(\ref{Eq:3pf}), in general, has two  different  contributions: i) One in which the insertion operator couples directly to a valence quark in the nucleon, leading to the so-called \emph{connected} three-point function, and ii)  one in which the current couples to a sea quark giving the \emph{disconnected} three-point function. In the case of the  flavor isovector  or octet currents, the disconnected contribution vanishes in the SU(3) flavor symmetric mass point and in the continuum limit. 
In the case of the flavor octet given in Eq.~(\ref{Eq:AVcurrentOctet}), and flavor singlet given in  Eq.~(\ref{Eq:AVcurrentSinglet}),  disconnected contributions are non-zero. Since for the octet combination the disconnected contribution vanishes only in the SU(3) flavor symmetric limit, any non-vanishing contribution  can be used to assess the level of SU(3) symmetric breaking.
For the evaluation of the connected contributions we employ standard techniques, as discussed in Ref.~\cite{Alexandrou:2020okk}, where we also give the statistics used for computing the connected contributions. 

Here we describe our approach to compute the  disconnected three-point functions. The disconnected quark loop for the axial-vector current  is given by
\begin{eqnarray}
L(t_{\rm ins},\vec{q}) = \sum_{\vec{x}_{\rm ins}} \tr \left[ D^{-1}(x_{\rm ins}; x_{\rm ins}) \gamma_\mu \gamma_5 \right] e^{+ i \vec{x} \cdot \vec{q}}.
\end{eqnarray}
The trace of the  all-to-all quark propagator $D^{-1}(x_{\rm ins}; x_{\rm ins})$, is the most computationally intensive quantity. Inverting from every point to every point on the lattice is computationally impossible for the lattice sizes
considered in this work and the resources available. Instead, we combine stochastic methods to estimate the value of the quark loop. A novel method we employ is the combination of   \emph{hierarchical probing}~\cite{Stathopoulos:2013aci} and deflation of low eigenvalues.  Hierarchical probing allows for  partitioning  the lattice up to a distance in a hierarchical manner using the Hadamard vectors as a basis. The partitioning is done through a coloring approach, up to a distance $2^k$ where $2^{d(k-1)+1}$ Hadamard vectors are needed ($d=4$ for a 4-dimensional coloring).  The computational cost of the method increases as $2^4$ each time one increases the coloring distance. Thus, even if the probing is done in a hierarchical manner allowing to reuse the results from previous distances, the gain is small when we increase further the distance. Contributions  from points beyond the probing distance are expected to be suppressed since the quark propagator decays exponentially fast with the distance from the diagonal. We further suppress such contributions  using stochastic vectors that have the properties
\begin{equation}
  \frac{1}{N_r} \sum_r \vert \xi_r \rangle \langle \xi_r \vert = \mathds{1} + \mathcal{O} \left( \frac{1}{\sqrt{N_r}} \right),
\end{equation}
and
\begin{equation}
    \frac{1}{N_r} \sum_r \vert \xi_r \rangle = 0,
\end{equation}
where $N_r$ is the number of stochastic vectors. The off-diagonal contributions are suppressed by $1/\sqrt{N_r}$. The hierarchical probing method was first employed in studies for heavier than physical pion masses~\cite{Green:2017keo,Green:2015wqa,Djukanovic:2019jtp} yielding results with unprecedented  accuracy.  For simulations at the physical point, it was shown~\cite{Alexandrou:2018sjm} that a  larger probing distance is required, as expected, since the light quark propagator decays slower due to the smaller quark mass. Instead of increasing the probing distance, which translates to a significant increase in computational cost, we combine hierarchical probing with deflation of the
low modes~\cite{Gambhir:2016uwp}. Namely, for the light quarks we construct
the low mode contribution to the  quark loops by computing exactly the smallest eigenvalues and corresponding eigenvectors of the squared Dirac operator and combine them with the contribution
from the remaining higher modes, which is estimated using hierarchical probing. Additionally, we fully dilute in spin and color and employ the \emph{one-end trick}~\cite{McNeile:2006bz}, that was employed in our previous studies~\cite{Alexandrou:2013wca,Alexandrou:2017hac,Alexandrou:2017oeh}.

The parameters used for the evaluation of the quark loops are collected in Table~\ref{table:StatsDisc}.  Two hundred  low modes of the square Dirac operator are computed in order to reduce the stochastic noise in the computation of the light quark loops. For the charm quark we use a coloring distance $2^2$ in hierarchical probing as compared to  $2^3$ used for the light and the strange quark loops. To increase the accuracy in the charm quark case we  compute 12 stochastic vectors instead of one used for the light and strange quark loops.
Nucleon two-point functions are evaluated  for two hundred randomly chosen source positions that are sufficient for reducing the gauge noise   for the large  sink-source time separations of the disconnected three-point functions. Since they are available,  we  use the same number of two point functions for all  sink-source time separations. 
  
\begin{table}[ht!]
  \caption{Parameters and statistics used  for the evaluation of the  disconnected quark loops for the cB211.072.64 ensemble.
    The number of configurations analyzed is $N_{\rm cnfs}=750$ and the number of source positions used for the evaluation of the two-point functions is $N_{\rm srcs}=200$ per gauge configuration. In the case of the light quarks, we compute the lowest  200 modes exactly
   and deflate before computing the higher modes stochastically. 
    $N_r$ is the number of noise vectors, and
    $N_{\rm Had}$ the number of Hadamard vectors. $N_{\rm sc}=12$ corresponds to spin-color dilution and
    $N_{\rm inv}$ is the total number of inversions per configuration.}
  \label{table:StatsDisc}
  \begin{center}
  \scalebox{1}{
  \begin{tabular}{c|c|c|c|c|c}
  \hline
    \hline
    Flavor & $N_{\rm def}$ & $N_r$ & $N_{\rm Had}$ & $N_{\rm sc}$ &  $N_{\rm  inv}$  \\
    \hline
     light & 200 & 1 & 512 & 12 & 6144   \\
      strange &0 & 1 & 512 & 12 & 6144    \\
      charm &0 & 12 & 32 & 12 & 4608   \\
    \hline\hline
  \end{tabular}
    }
  \end{center}
\end{table}

For disconnected quantities we are not limited to using $\vec{p}\,'=0$ since no additional inversions are needed.  Therefore, we consider several values
of $\vec{p}\,'$, namely $\vec{p}\,'=(2\pi/L)\, \vec{n}\,'$ up to $\vec{n}^{\prime 2}=2$  and for $\vec{p}$ up to $\vec{n}^2=22$. This allows us to compute the disconnected parts of the form factors for a higher density of $Q^2$ values.

\section{Renormalization functions}\label{sec:Renormalization}
In order to relate the matrix elements computed on the lattice to physical observables one needs to renormalize. Here, we summarize our procedure. A more detailed description can be found in Ref.~\cite{Alexandrou:2015sea}. We employ a mass-independent renormalization scheme  and analyze five $N_f=4$ ensembles generated specifically for the determination of the renormalization functions. The value of $\beta$ is the same as that of the cB211.072.64 ensemble. The pion masses are in the range of [366-519]~MeV. These are used to take the chiral limit. The lattice volume is $24^3\times48$ for all $N_f=4$ gauge ensembles. The Rome-Southampton method, RI$^\prime$ scheme~\cite{Martinelli:1994ty}, is employed where the quark propagators and vertex functions are non-perturbatively determined. For the axial-vector operator we need to renormalize with $Z_A$ and, since we consider also disconnected contributions, both singlet and non-singlet renormalization factors are needed.

We impose the following renormalization conditions: 
\begin{equation}
    Z_q = \frac{1}{12} \tr \left[ (S^L(p))^{-1} S^{\rm Born}(p) \right]\bigg\vert_{p^2=\mu_0^2}
\end{equation}
and
\begin{equation}
\label{eq:RI_cond}
    Z_q^{-1} Z_{\cal O}\frac{1}{12} \tr \left[ (\Gamma^L(p)) \Gamma^{\rm Born-1}(p) \right]\bigg\vert_{p^2=\mu_0^2}=1.
\end{equation}
 $S^L(p)$ and $\Gamma^L(p)$ are the quark propagator and amputated vertex function, respectively, while $S^{\rm Born}(p)$ and $\Gamma^{\rm Born}(p)$ are the corresponding tree-level values. We note that the trace is meant to be taken over both spin and color indices and the RI$^\prime$ renormalization scale is denoted by $\mu_0$.
In order to compute the vertex functions non-perturbatively, we make use of momentum sources~\cite{Gockeler:1998ye}. This allows to achieve per mil statistical accuracy on a very small sample of configurations~\cite{Alexandrou:2010me,Alexandrou:2012mt}. With a high statistical precision, systematic errors need to also be under control. The momenta are chosen isotropic in the spatial direction, that is
\begin{equation}
  (a\,p) \equiv 2\pi \left(\frac{2n_t+1}{2T/a},
\frac{n_x}{L/a},\frac{n_x}{L/a},\frac{n_x}{L/a}\right),
\end{equation}
where $n_t\in[2, 10],\,n_x\in[2, 5]$ and $T/a$($L/a$) are the temporal(spatial) lattice extent. The momenta satisfy the condition ${\sum_i p_i^4}/{(\sum_i p_i^2)^2}{<}0.3$~\cite{Constantinou:2010gr} in order to suppress the non-Lorentz invariant contributions. These appear in ${{\cal O}(a^2)}$ terms in the perturbative expansion of the Green's function and is expected to have non-negligible contributions from higher order in perturbation theory~\cite{Alexandrou:2010me,Alexandrou:2012mt,Alexandrou:2015sea}. 

We improve the non-perturbative estimates by removing lattice artifacts in both $Z_q$ an $Z_A$. The artifacts are calculated to one loop lattice perturbation theory~\cite{Alexandrou:2015sea}. In particular, one extracts the Greens functions of the axial operator using the same lattice action and values of the momentum $p$ entering Eq.~(\ref{eq:RI_cond}). For an optimal improvement, we calculate ${\cal O}(g^2 a^\infty)$ terms, which cannot be obtained analytically. It should be noted that, the subtraction of the ${\cal O}(g^2 a^\infty)$ terms can be done either at the level of the vertex functions $\Gamma_L(p)$ , or on $Z_A$ after the trace is taken. We have checked that both procedures lead to compatible results for the improved $Z_A$. For consistency, we employ the subtraction in the final estimates of $Z_A$, as performed in Ref.~\cite{Alexandrou:2020okk}.

The evaluation of the Z-factors for the non-singlet current was presented in Ref.~\cite{Alexandrou:2020okk}. Here we  present the evaluation of the singlet Z-factor,  which is more complicated. 
For the computation of the singlet renormalization function $Z_A^s$ we follow the same procedure as for the non-singlet case. In this case, in addition to the connected contributions, there are contributions from the disconnected quark loops. We employ the same noise reduction approaches discussed in Sec.~\ref{ssec:ConnDiscStats} for the evaluation of these disconnected contributions, namely we use hierarchical probing with 512 Hadamard vectors, the one-end trick and spin color dilution. Deflation is not  used in this case since the $N_f$ ensembles are generated for heavy pion masses. 
In addition to the appearance of disconnected loops, a further complication is that, in contrast to the non-singlet case,  $Z_A^s$   is scheme and scale dependent. We express it in the ${\overline{\rm MS}}$-scheme, which is commonly used in experimental and phenomenological studies. The conversion procedure is applied on the Z-factors obtained on each initial RI$'$ scale $(a\,\mu_0)$, with a simultaneous evolution to a $\overline{\rm MS}$ scale, chosen to be $\overline{\mu}{=}$2 GeV. In particular, we use the conversion factor calculated to 2-loops in
perturbation theory~\cite{Skouroupathis:2008mf}. 

\begin{figure}[ht!]
 \includegraphics[scale=0.4]{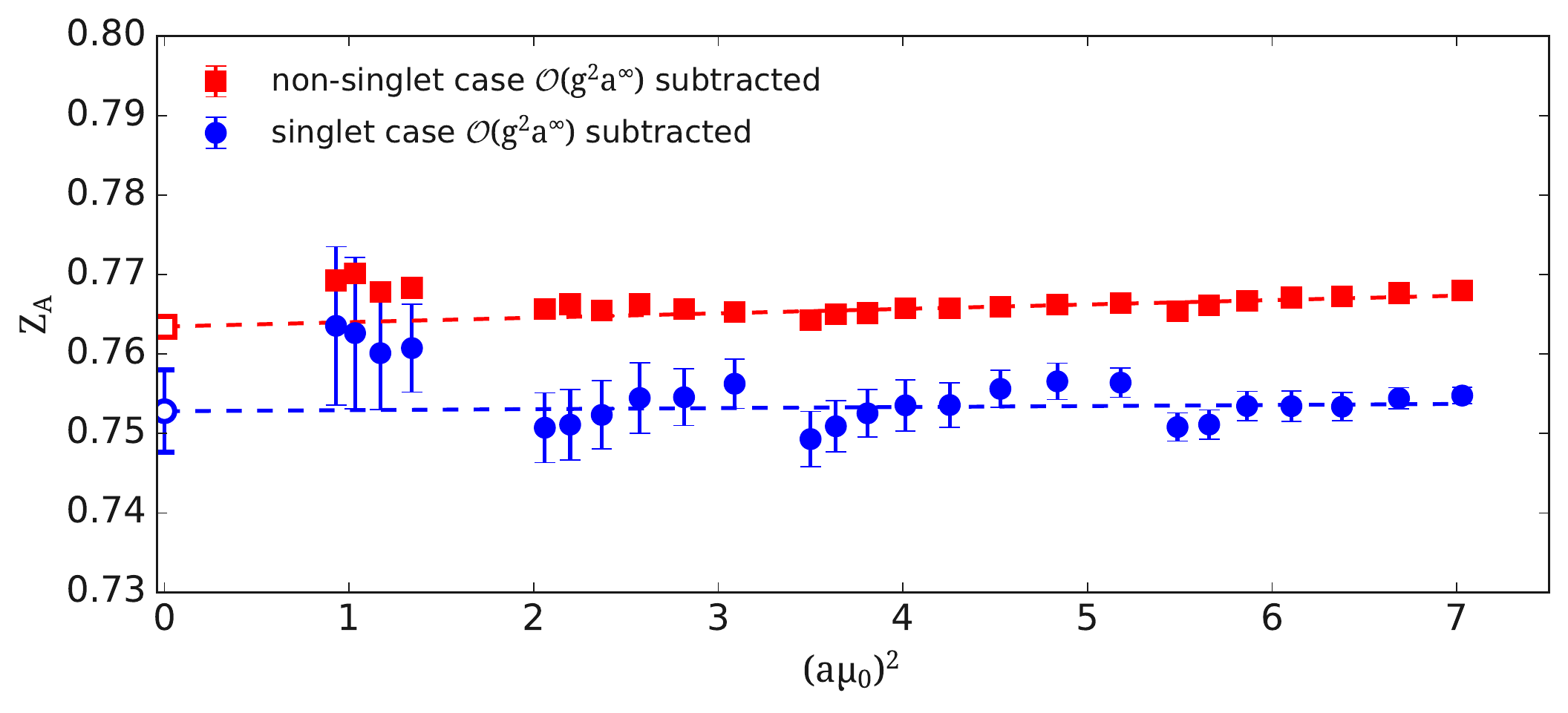}
 \caption{Results for the non-singlet (red squares) and singlet (blue cicles) $Z_A$ as a function of the initial renormalization scale $(a\mu_0)^2$. The dashed lines are linear fits and  open symbols are the extracted extrapolated values.}
 \label{fig:ZAs_vs_ap2}
 \end{figure}

 In Fig.~\ref{fig:ZAs_vs_ap2} we compare the non-singlet and singlet Z-factors. As can be seen, including the disconnected quark loop contributions lowers  the value of the renormalization function and increases  the error. We find $Z_A=0.763(1)$ for non-singlet and $Z_A^s=0.753(5)$ for singlet.

\section{Analysis of the isoscalar axial form factors $G_A^{u+d}(Q^2)$ and $G_P^{u+d}(Q^2)$}
\label{sec:isos}

\begin{widetext}
\begin{figure*}[ht!]
 \includegraphics[scale=0.55]{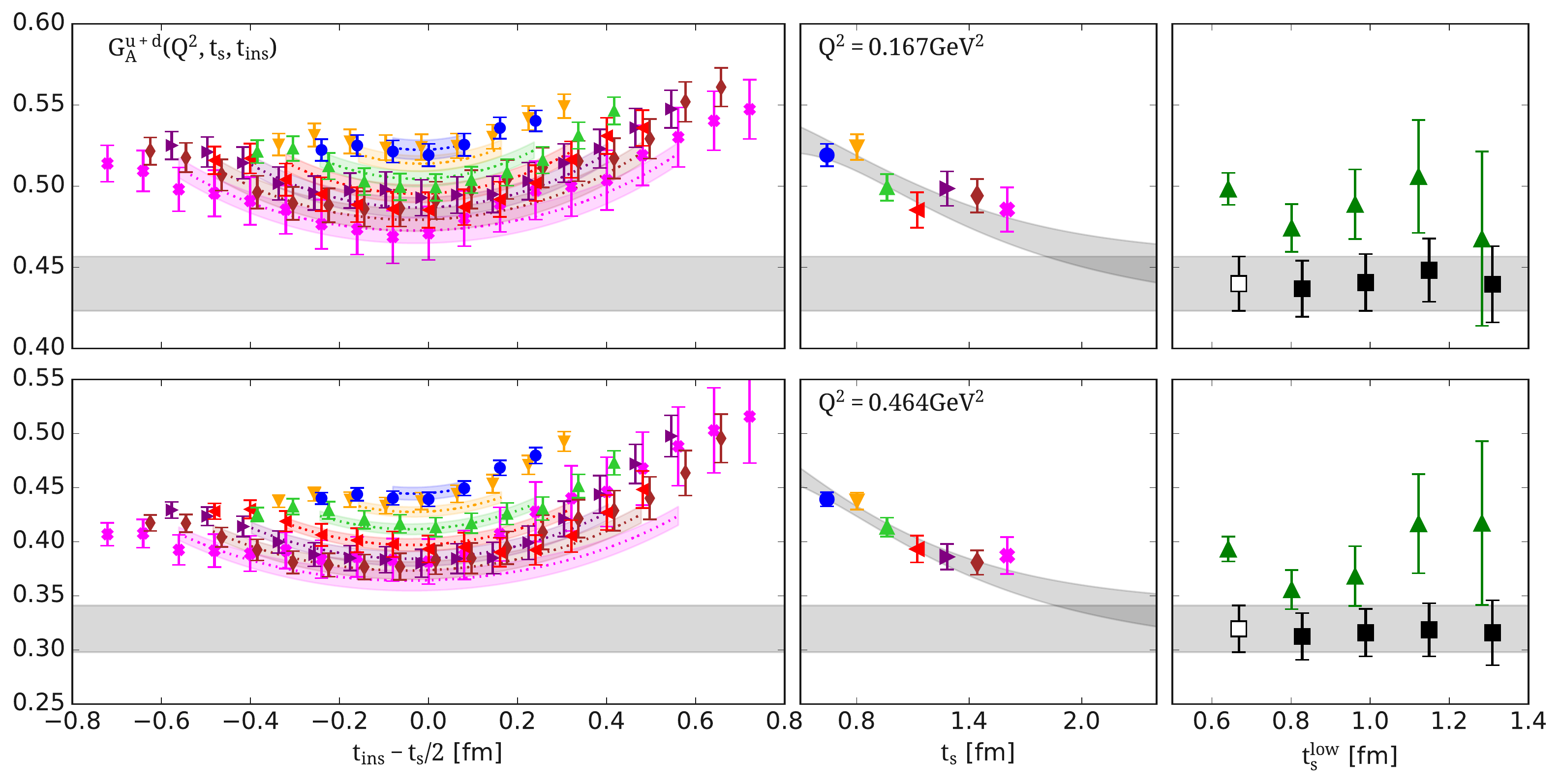}
 \caption{
 Results on the connected renormalized $G_A^{u+d}(Q^2)$ extracted using  the plateau, two-state and the summation methods  for two different values of $Q^2$, namely for  $Q^2=0.167$~GeV$^2$ (top row) and $Q^2=0.464$~GeV$^2$ (bottom row).
 In the left panel, we show results on the ratio of Eq.~(\ref{Eq:ratio}) for sink-source time separations $t_s/a=8,10,12,14,16,18,20$ denoted with blue circles, orange down triangles, up green triangles, left red triangles, right purple triangles, brown rhombus and magenta crosses, respectively. The results are shown as a function of the insertion time $t_{\rm ins}$ shifted by $t_s/2$. The dotted lines and  associated error bands  are the resulting 
  two-state fits when the lowest value of $t_s$ used in the fit ($t_s^{\rm low}$) is
   $t_s^{\rm low}=8a=0.64$~fm. In the middle panel, we show  the plateau values or the value of the ratio for $t_{\rm ins} = t_s/2$ when no plateau is identified, as a function of $t_s$ using the same  symbol for each  $t_s$ as used for the ratio in the left panel. In the right panel, we show the extracted values using the two-state fit (black squares) and the summation method (green filled triangles) as a function of $t_s^{\rm low}$. The open symbol shows our selected value with the grey band spanning the whole range of the figure being the associated statistical error.  The color bands on the left column are the predicted time-dependence of the ratio using the parameters extracted  from the two-state fit when  $t_s^{\rm low}=8a=0.64$~fm.  The $\chi^2/$d.o.f is 1.09 for $Q^2=0.167$~GeV$^2$ and 1.24 for $Q^2=0.167$~GeV$^2$.} 
 \label{fig:GA_isos_effFFs}
 \end{figure*}
\end{widetext}

For the extraction of the axial and induced pseudoscalar form factors from the correlation functions we use the methodology presented in Sec.~\ref{sec:LatM}. In order to  identify the nucleon ground-state contribution  we apply the three approaches  discussed in Sec.~\ref{ssec:ExcStates}.  In  Fig.~\ref{fig:GA_isos_effFFs}, we demonstrate the effect of the excited-states contamination to the connected contribution for the isoscalar axial form factor  $G_A^{u+d}(Q^2)$ for two representative values of $Q^2$. 
In the first column, we show the ratio of Eq.~(\ref{Eq:ratio}) for all the available values of $t_s$.  In the construction of the ratio, we use the two-point functions computed with the same source positions as the corresponding three-point functions to exploit their correlation leading to a reduction in the overall error.
As $t_s$ increases, we observe a decrease in the values of the ratio. In the second column of Fig.~\ref{fig:GA_isos_effFFs} we show the values extracted by fitting the ratio to a constant excluding five time slices from the source and sink. This is done for  $t_s/a > 12$, yield a good $\chi^2$/d.o.f, namely in the range 0.7 to 1.1.
In the third column of Fig.~\ref{fig:GA_isos_effFFs} we show results from the two-state and summation fits. For the two-state method we perform a simultaneous  fit to all ratios for which $t_s \ge t_s^{\rm low}$ excluding $t_{\rm ins}/a=1,2, t_s-1,t_s-2$ and seek to identify convergence in the extracted value of the matrix element as we increase $t_s^{\rm low}$. The resulting fit bands using these two-state fits are shown in the left panel. We show the prediction of the two-state fit of the time dependence of the ratio in the middle panel when we fix $t_{\rm ins}=t_s/2$. As can be seen, the two-state fit prediction describes well the time dependence of the values extracted from the plateau method.
We also observe that the value extracted using the two-state fit with $t_s^{\rm low}=8a$ is consistent with the values extracted for higher values of $t_s^{\rm low}$. The results from the  summation method converge to those of the two-state fit for $t_s^{\rm low}> 1.2$~fm  i.e. at about half the $t_s$ value where the plateau fit yields convergent result.
Based on these findings, we adopt as a criterion for the final value the one extracted from the two-state fit for the smallest $t_s^{\rm low}$ that shows convergence and is in agreement with the value from the summation method at some higher $t_s^{\rm low}$. Our final value  is indicated with the open symbol in Fig.~\ref{fig:GA_isos_effFFs}.

\begin{widetext}
 \begin{figure*}[h!]
 \includegraphics[scale=0.525]{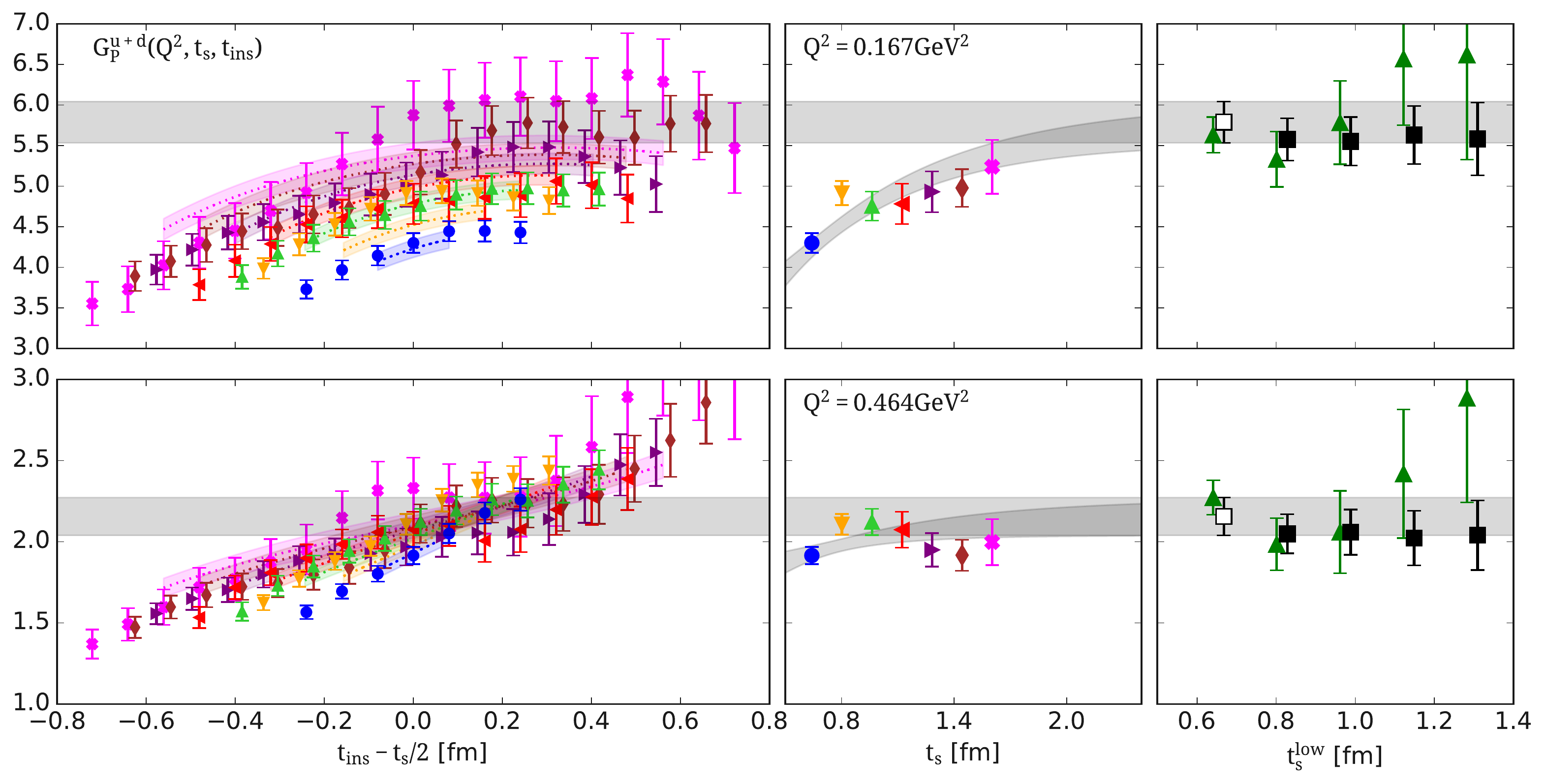}
 \caption{Results for the connected renormalized induced pseudoscalar form factor $G_P(Q^2)$. The notation is the same as that  in Fig.~\ref{fig:GA_isos_effFFs}. } 
 \label{fig:GP_isos_effFFs}
 \end{figure*}
 \end{widetext}

In Fig.~\ref{fig:GP_isos_effFFs}, we present the excited-states contamination analysis for the case of the connected contributions to the isoscalar induced pseudoscalar form factor. In contrast to $G_A$, suppression of  excited states results in larger values for $G_P(Q^2)$ especially for the smaller $Q^2$ values.  As $Q^2$ increases, contamination from excited states suppresses,
 with most of the plateau values being compatible with the two-state fit. We use the same criterion as for $G_A^{u+d}$ for the selection of our final values. Therefore, we take the values extracted from the two-state fit  for $t_s^{\rm low}=8a$.

 \begin{widetext}
\begin{figure*}[!ht]
     \centering
     \includegraphics[scale=0.525]{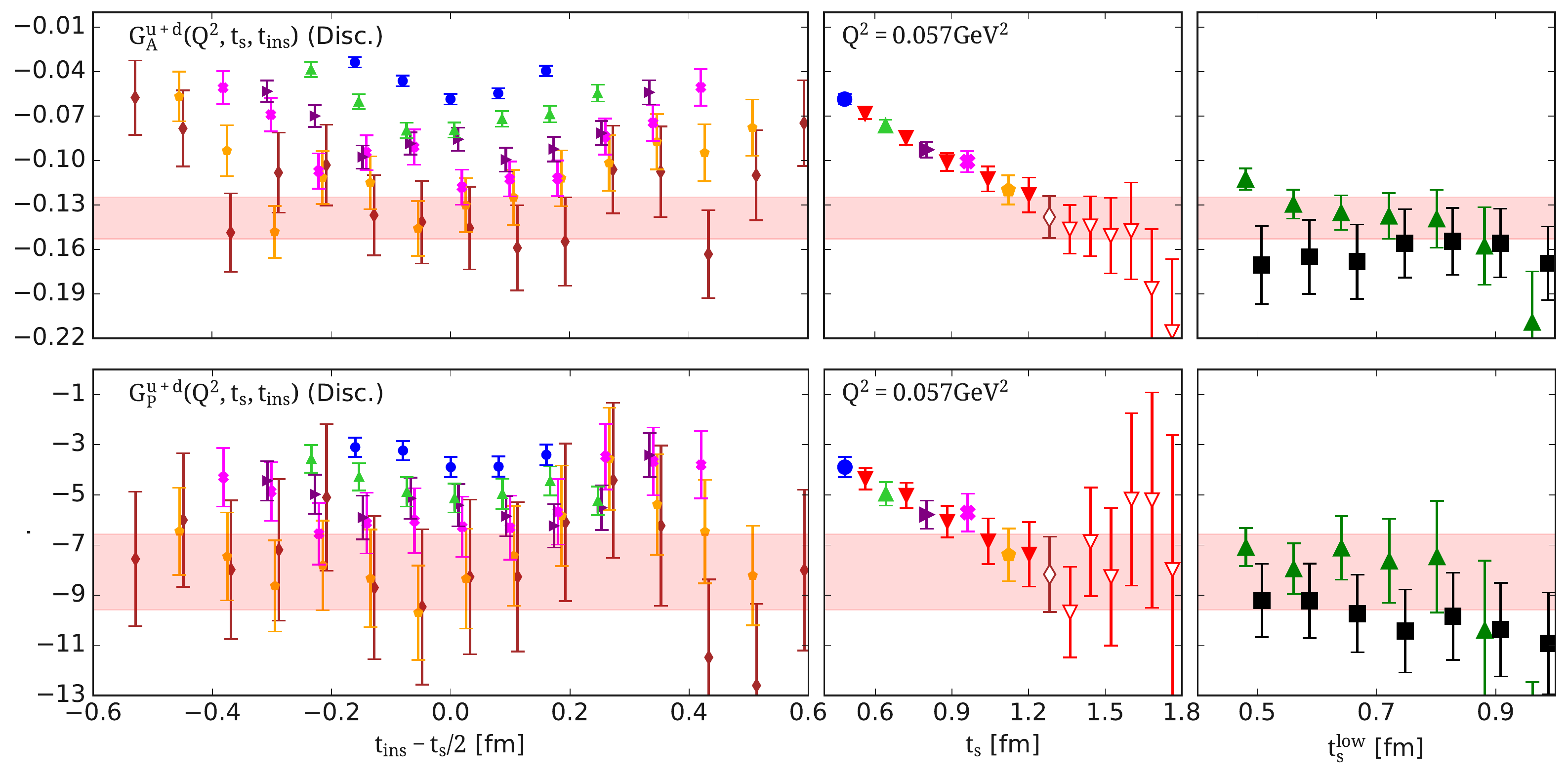}
     \caption{Results  on the renormalized disconnected parts of $G_A^{u+d}(Q^2)$ (top) and $G_{P}^{u+d}$ (bottom) for $Q^2=0.057$~GeV$^2$ extracted using the plateau, two-state fit and the summation methods. In the left panel we show results on the ratio of Eq.~(\ref{Eq:ratio}) for sink-source time separations $t_s/a=6,8,10,12,14,16$ denoted with blue circles, up green triangles, right purple triangles, magenta crosses, orange pentagons, and brown rhombus, respectively. For the middle column we show the values extracted from the plateau fits using the same color and symbol as for the corresponding $t_s$ shown in the left column. In the constant fits we exclude $t_{\rm ins}/a=1,2, t_s-1,t_s-2$ points for each $t_s$. The red downward triangles show the values for intermediate (odd values $t_s/a$) or larger $t_s$ not drawn for clarity in the left panel. The horizontal red band denotes the final value computed as a weighted average using the converged plateau values indicated with open symbols. The notation on the right panel is the same as that in Fig.~\ref{fig:GA_isos_effFFs}.}
     \label{fig:GA_Gp_effFFs_light}
 \end{figure*}
\end{widetext}
 
In  Fig.~\ref{fig:GA_Gp_effFFs_light}, we show the analysis to identify excited-state contributions for the disconnected parts contributing to $G_A^{u+d}(Q^2)$ and $G_P^{u+d}(Q^2)$. Although in these cases, all the sink-source time separations can be computed without additional cost, in practice, as the time separation $t_s$ increases, the errors become very large. Thus, we limit ourselves to $t_s \in [0.48-1.8]$~fm in what follows.  As can be seen for both form factors, the disconnected contributions are non-zero. Eliminating excited states by increasing $t_s$ leads to more negative values for both axial and induced pseudoscalar form factors.
In both cases, results for $t_s\ge16a$ extracted from the plateau method are in agreement with each other, as well as with those extracted using the two-state and summation methods. We thus opt to perform a weighted average of the converged plateau values to extract the final value.

\begin{figure}[h!]
 \includegraphics[scale=0.525]{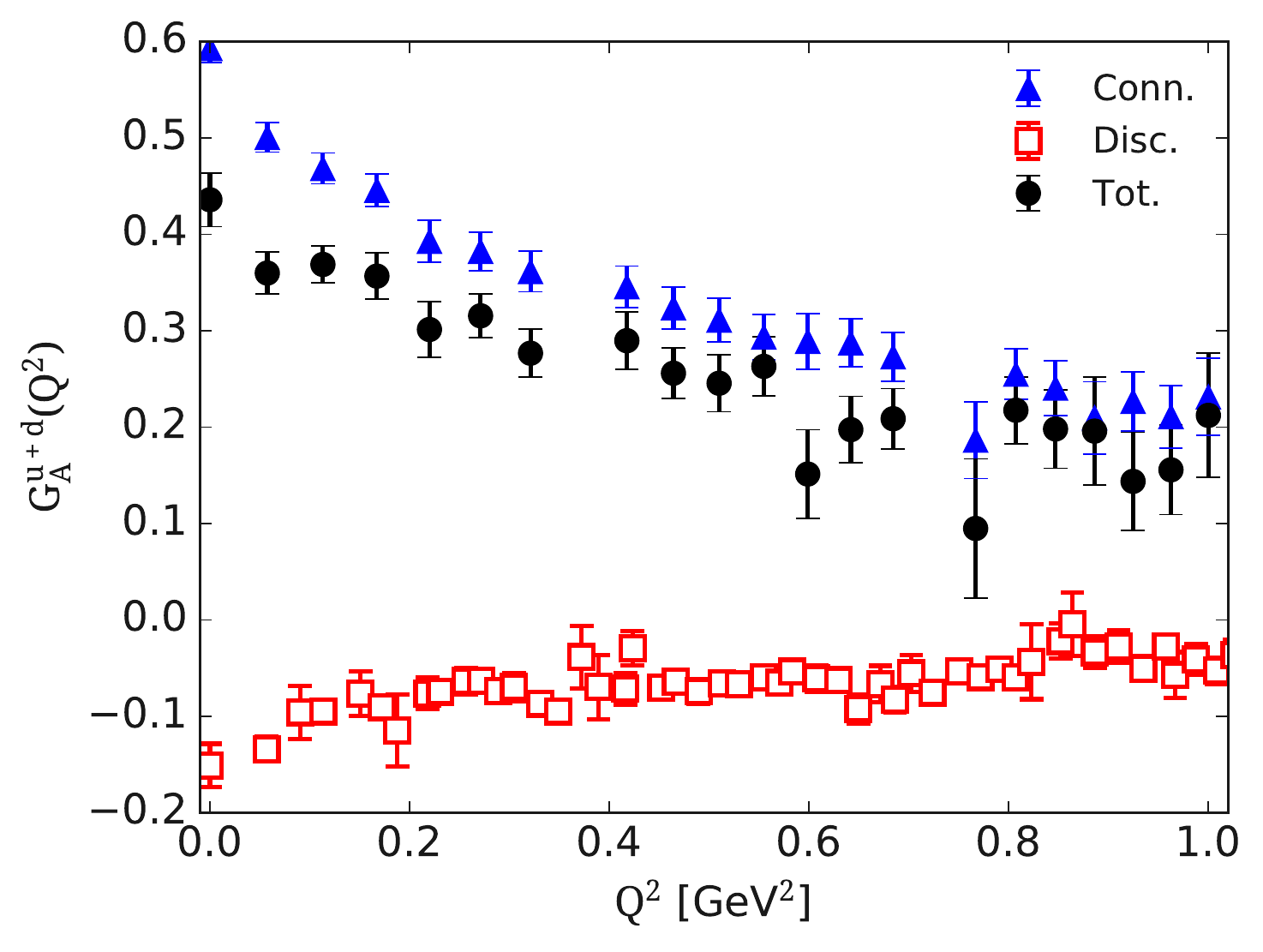}
 \includegraphics[scale=0.525]{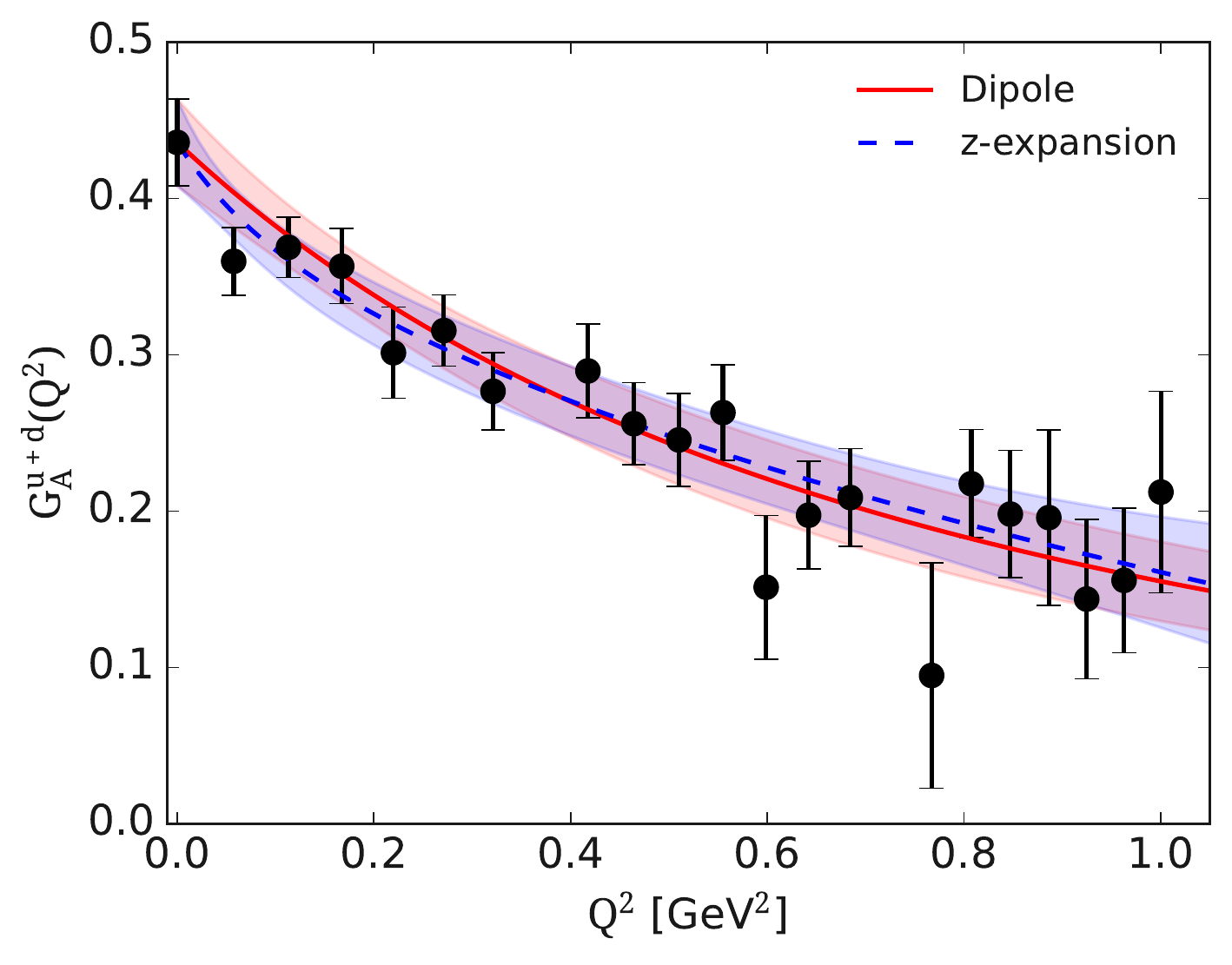}
 \caption{\emph{Left:} Renormalized results for $G_A^{u+d}(Q^2)$ as a function of  $Q^2$. We show separately the connected   (blue triangles) and the disconnected (open red squares) contributions as well as the sum (black circles). Open symbols are used for the form factors versus $Q^2$ when showing only disconnected contributions. \emph{Right:} Renormalized results for $G_A^{u+d}(Q^2)$ as a function of $Q^2$.   The solid red line is the result of the dipole fit and the dashed blue of the z-expansion fit. The red and blue bands are associated with the uncertainties of the dipole and z-expansion fits. Note that the upper fit range is  1~GeV$^2$. }
 \label{fig:GA_isos_f1}
 \end{figure}

The final values of the axial form factor $G_A^{u+d}(Q^2)$ are shown in the left panel of Fig.~\ref{fig:GA_isos_f1}, where we show separately the connected and disconnected contributions as a function of $Q^2$. We observe  that  the connected contribution is positive, while the disconnected is  negative. To extract the disconnected part, we combine various values of  $\vec{p}\,' \ge \vec{0}$ and thus can access a larger number of $Q^2$ values.  We  show also $G_A^{u+d}(Q^2)$  after summing the connected and disconnected parts at the common $Q^2$ values.  Since the disconnected contributions have a larger magnitude at smaller $Q^2$ values the slope of $G_A^{u+d}(Q^2)$ at small $Q^2$ is smaller as compared  to its connected part.  We fit the $Q^2$ dependence as shown in the right panel of Fig.~\ref{fig:GA_isos_f1} using the dipole Ansatz and the z-expansion, as described in Sec.~\ref{sec:A-V_ME}, where for both fits the value at $Q^2=0$ is not a fit parameter but it is fixed form the forward matrix element yielding  $G_A^{u+d}(0) \equiv g_A^{u+d}$. We find $g_A^{u+d}=0.436(28)$ in agreement with our previous study~\cite{Alexandrou:2019brg}. The small difference is due to fact that in this work we use also $\vec{p}\,' > \vec{0}$ for the evaluation of the disconnected contributions, but also odd numbers of $t_s$ when averaging over the plateau values.  Both  fit forms describe the $Q^2$ behavior very well. The extracted values for the axial masses and the radii are given in Table~\ref{table:GA_isos_extr_vals}. The values  extracted from the two fits are compatible, with the z-expansion yielding larger uncertainties. We note that by excluding larger values of $Q^2$ in the fit does not have an impact on the extracted parameters.

\begin{widetext}
   \begin{table*}[h!]
  \caption{
  Extracted values from the fits on the isoscalar axial form factor as in the right panel of Fig.~\ref{fig:GA_isos_f1} using the dipole Ansatz and the z-expansion. We use two ranges for the largest $Q^2$ value included in the fit, one up to $Q^2\simeq0.5$~GeV$^2$ and the second up to $Q^2\simeq1$~GeV$^2$. The extracted parameters are the axial mass, $m_A$ and the root mean square  (r.m.s) radius, $\sqrt{\langle (r_A^{u+d})^2 \rangle}$. In the last column we give the $\chi^2$ per degrees of freedom (d.o.f). We make use of Eq.~(\ref{Eq:rToM}) to relate the mass to the radius and vice versa. The isoscalar charge,  $g_A^{u+d}=G_A^{u+d}(0)$ is 0.436(28).}
  \label{table:GA_isos_extr_vals}
  \vspace{0.2cm}
  \begin{tabular}{c|c|c|c|c|c}
    \hline
    Fit Type & $Q^2_{\rm max}$~[GeV$^2$] & $m_A^{u+d}$ [GeV]  & $\sqrt{\langle (r_A^{u+d})^2 \rangle}$ [fm] & $\chi^2/$d.o.f \\
    \hline
    \multirow{2}{*}{Dipole} & $\simeq$ 0.5 & 1.188(169) &  0.575(82) & 0.79\\
    & $\simeq$ 1& 1.216(144) &  0.562(67) & 0.72 \\
    \hline
    \multirow{2}{*}{z-expansion} & $\simeq$ 0.5 & 0.949(215) &  0.720(163) & 0.49\\
    & $\simeq$ 1& 0.975(234) &  0.701(168) & 0.59 \\
    \hline\hline
  \end{tabular}
\end{table*}
 \end{widetext}
 
\begin{figure}[ht!]
 \includegraphics[scale=0.525]{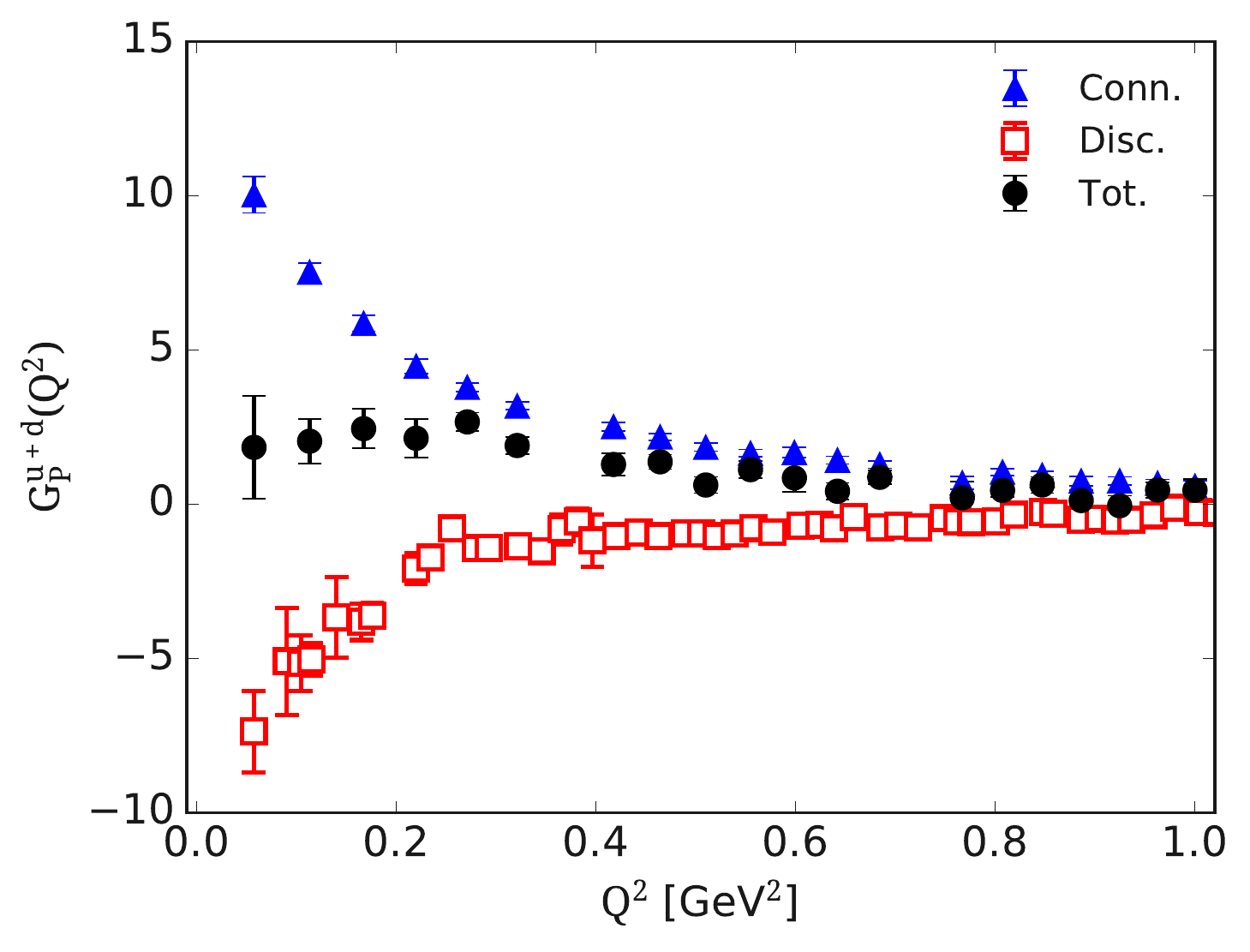}
 \caption{Renormalized results for $G_P^{u+d}(Q^2)$ as a function of  $Q^2$. The notation is as in the left panel of Fig.~\ref{fig:GA_isos_f1}. }
 \label{fig:GP_isos_f1}
 \end{figure}

 In Fig.~\ref{fig:GP_isos_f1} we show separately the connected and disconnected parts for the isoscalar induced pseudoscalar form factor $G_P^{u+d}(Q^2)$. The disconnected part is of the same magnitude as the connected but with opposite sign. This has already been observed in previous studies~\cite{Green:2017keo,Alexandrou:2017hac}. This behavior leads to the cancellation of the sharp rise observed in the connected $G_P^{u+d}(Q^2)$. Consequently, the isoscalar has an almost flat $Q^2$-dependence within uncertainties, unlike the isovector combination where the pion pole gives a rapidly rising form factor at small $Q^2$~\cite{Alexandrou:2020okk}. Also, the fact that the connected and disconnected parts are almost equal but with opposite sign means that $G_P^{u+d}(Q^2)$ carries larger statistical errors. 
 
 \vspace*{0.5cm}
 
 \section{Analysis of the strange and charm axial form factors} \label{sec:SC_FFs}
\begin{widetext}
\begin{figure*}[!h]
     \centering
     \includegraphics[scale=0.525]{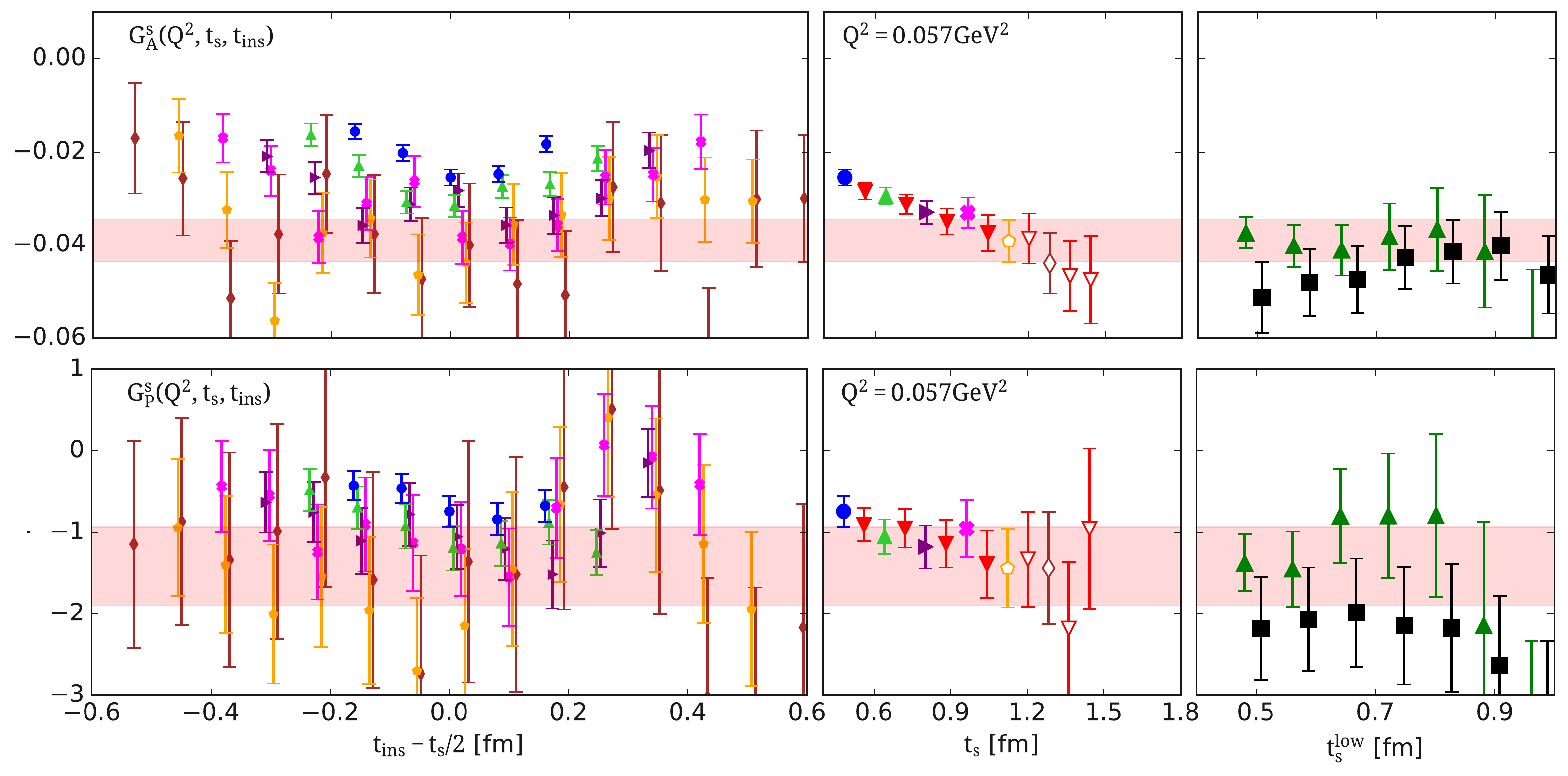}
     \caption{Results on the renormalized strange axial form factors $G_A^{s}(Q^2)$ (top)   and $G_{P}^{s}$ (bottom) for $Q^2=0.057$~GeV$^2$ extracted using the plateau, two-state fit and the summation methods. In the middle panels, open symbols denote the plateau values that we take into account in the weighted average resulting in our final value shown with  the red band. The rest of the notation is the same as that in Fig.~\ref{fig:GA_Gp_effFFs_light}.}
     \label{fig:GA_Gp_effFFs_strange}
 \end{figure*}
\end{widetext}

The strange and charm form factors receive only purely disconnected contributions. They probe sea quark degrees of freedom in the nucleon  and provide us with an insight on their non-perturbative dynamics. Let us first examine how the ratio of Eq.~(\ref{Eq:ratio}) behaves when using the three-point function of the strange axial-vector current.
 In Fig.~\ref{fig:GA_Gp_effFFs_strange}, we show the results on the ratio for different $t_s$. As can be seen, although there is a trend to more negative values, the plateau region is consistent within the statistical uncertainties as we increase $t_s$. This is also seen in the middle panel where we show the values extracted from plateau fits at various $t_s$ values.
 Furthermore, the summation and two-state fit methods yield results that are consistent with those extracted from the plateau fit for all  $t_s^{\rm low}$ values.  Given that the plateau values show  convergence, we take the weighted  average over the converged plateau values observed for $t_s \simeq 1.12$~fm, resulting in the red band.  The weighted average is also in agreement with the results from the two-state and summation fits, as we require to accept the final value.

\begin{widetext}
\begin{figure*}[!h]
     \centering
     \includegraphics[scale=0.525]{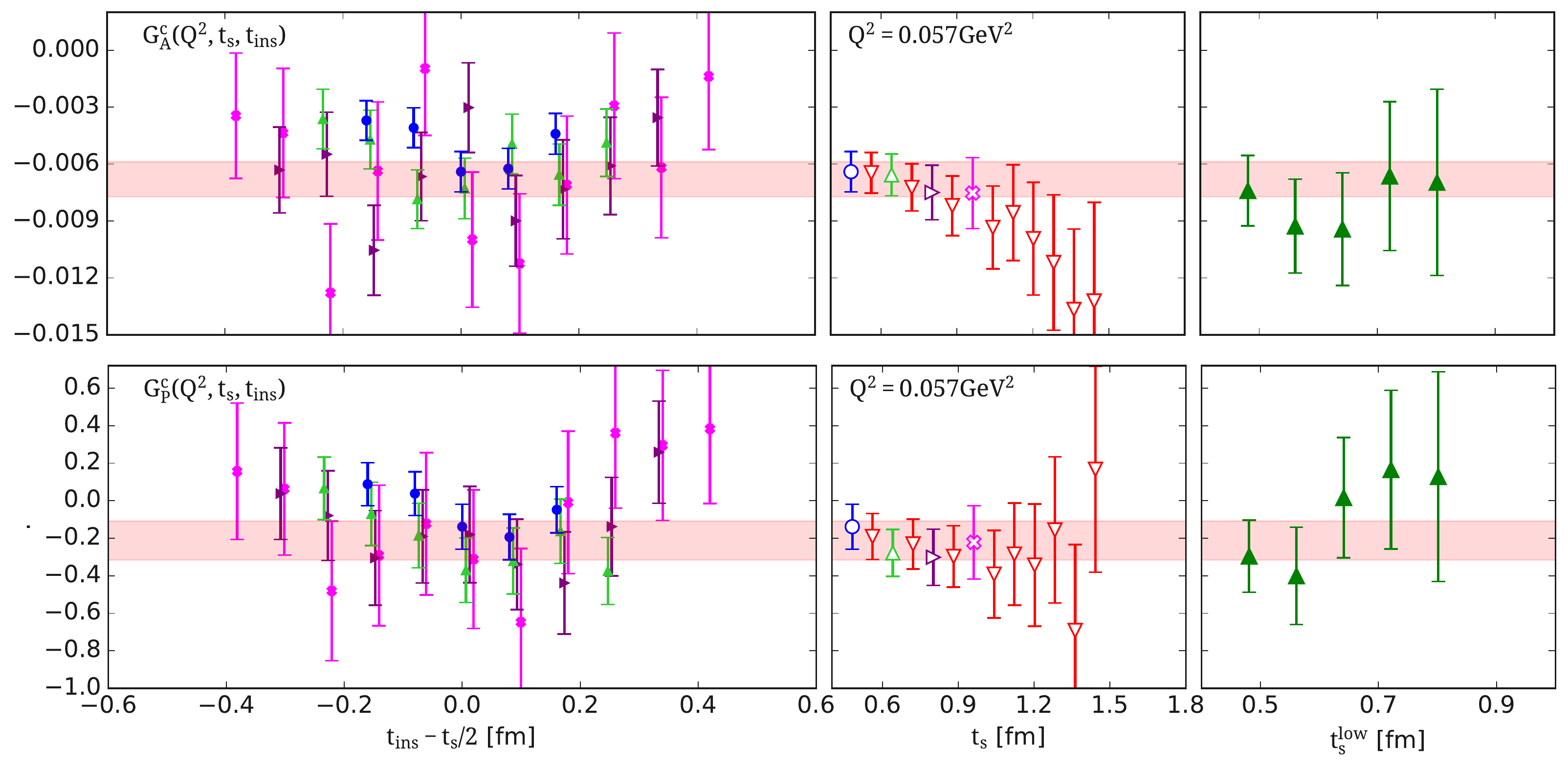}
     \caption{Results on the renormalized charm axial form factors $G_A^{c}(Q^2)$ (top)   and $G_{P}^{c}$ (bottom) for  $Q^2=0.057$~GeV$^2$ extracted using the plateau and the summation methods. 
     The notation is the same as in Fig.~\ref{fig:GA_Gp_effFFs_light}. }
     \label{fig:GA_Gp_effFFs_charm}
 \end{figure*}
\end{widetext}

The corresponding analysis of excited states for the three-point function of the charm axial-vector current  is shown in Fig.~\ref{fig:GA_Gp_effFFs_charm}. The three-point function in this case is  more noisy and for clarity we only show the  ratio for time separations up to 1~fm. As in the case of the strange three-point function,
the plateau region of the ratio shows convergence as $t_s$ is increased within our current statistical accuracy. The results extracted using the summation method are noisy but yield  consistent values.  Two-state fits are omitted since,  given the accuracy of the data, they are very noisy and thus yield no useful information. We  take the weighted average of the converged plateau values to determine the final values on $G_A^c(Q^2)$ and $G_P^c(Q^2)$.

\begin{figure}[!ht]
     \centering
     \includegraphics[scale=0.525]{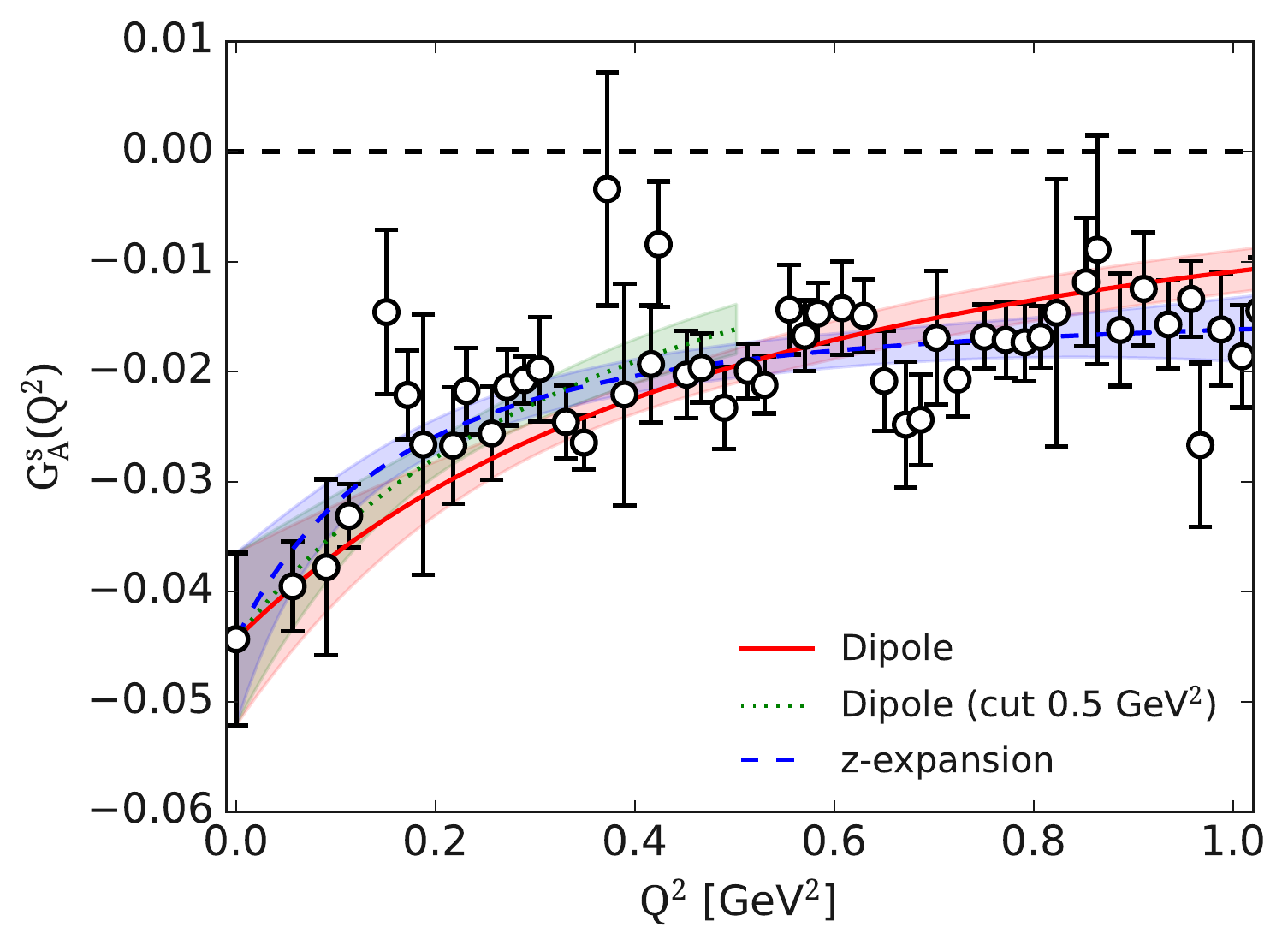}
     \includegraphics[scale=0.525]{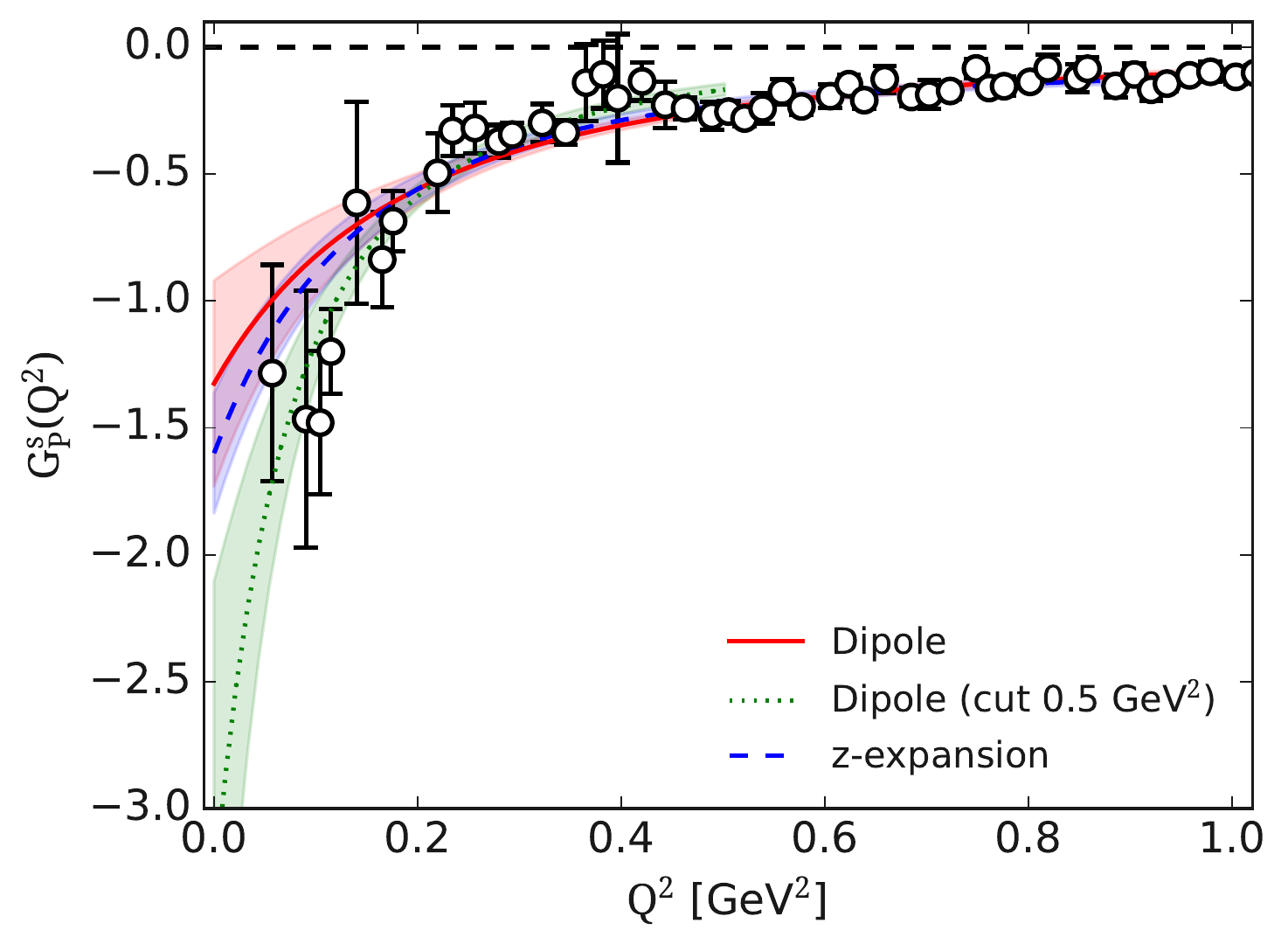}
     \caption{ Results for the strange form factor $G_A^s(Q^2)$ (left) and  $G_P^s(Q^2)$ (right) as a function of $Q^2$. Following the notation of Fig.~\ref{fig:GA_isos_f1}, we use open symbols when plotting the form factors as a function of $Q^2$ when only disconnected contributions enter. We also show the fit  using the dipole form taking the upper  fit range up to $\simeq$0.5~GeV$^2$ (green dotted line and band). The rest of the notation is the same as in Fig.~\ref{fig:GA_isos_f1}.}
     \label{fig:GAGP_strange}
 \end{figure}

The results for the strange axial form factor $G_A^s(Q^2)$ are shown in left panel of Fig.~\ref{fig:GAGP_strange}. $G_A^s(0)$ gives the strange axial charge and we find $g_A^s=-0.044(8)$ 
in agreement with the values reported in our previous analysis using the cB211.072.64 ensemble~\cite{Alexandrou:2019brg}. The small difference in the mean value is well within errors and is  due to taking different data sets in the analysis. $G_A^s(Q^2)$ is negative for all $Q^2$ values up to 1~GeV$^2$. Both fits to a dipole form and the z-expansion describe the data well. The value at $Q^2=0$ is used as an input parameter. In Table~\ref{table:GA_s}, we give the  $\chi^2/$d.o.f for the fits. The reason for the smaller $\chi^2/$d.o.f for the z-expansion is that higher order terms are taken into account that are sensitive to the values at larger $Q^2$ values giving rise  to more curvature and thus a somewhat better description of the data. 

The extracted values for  the strange axial mass $m_A^s$ and r.m.s. radius  $\sqrt{\langle (r_A^s)^2 \rangle}$ are  given in Table~\ref{table:GA_s}.  Although the  z-expansion fit has a steeper slope as $Q^2\rightarrow 0$ as compared to the dipole fit (see left panel of Fig.~\ref{fig:GAGP_strange}) the resulting values of the  radius are  consistent within the uncertainties.  
The strange induced pseudoscalar form factor $G_P^s(Q^2)$ as a function of $Q^2$ is shown in the right panel of Fig.~\ref{fig:GAGP_strange}. As in the case of  $G_A^s(Q^2)$, $G_P^s(Q^2)$ is clearly negative and large in magnitude especially at low $Q^2$.  The dipole and the z-expansion fits describe well the data. 
However, when we limit the fit range  up to $Q^2=0.5$~GeV$^2$ the r.m.s and the value of the form factor at $Q^2=0$ are significantly larger. 
This is due to the curvature observed for small $Q^2$.
\begin{widetext}
 \begin{table*}[h!]
  \caption{Parameters extracted from $G_A^s(Q^2)$ and $G_P^s(Q^2)$ using the dipole Ansatz and the z-expansion. The notation is the same as that in Table~\ref{table:GA_isos_extr_vals} up to column five. The next columns are $G_p^s(0)$, the value of the induced pseudoscalar form factor for $Q^2=0$, $m_P^s$ the dipole mass and $\sqrt{\langle (r_P^{s})^2 \rangle}$ the r.m.s radius.}
  \label{table:GA_s}
  \vspace{0.2cm}
  \begin{tabular}{c|c|c|c|c||c|c|c|c|c}
    \hline
    Fit Type & $Q^2_{\rm max}$~[GeV$^2$] & $m_A^s$ [GeV] &  $\sqrt{\langle (r_A^s)^2 \rangle}$ [fm] & $\chi^2$/d.o.f & $G_P^s(0)$ & $m_P^s$ [GeV] &  $\sqrt{\langle (r_P^s)^2 \rangle}$ [fm] & $\chi^2$/d.o.f \\
    \hline
    \multirow{2}{*}{Dipole} & $\simeq$ 0.5 & 0.874(162) &  0.782(145) & 1.33 & -3.328(1.224) & 0.381(59) & 1.796(276) & 0.91\\
    & $\simeq$ 1& 0.992(164) &  0.689(114) & 1.48 & -1.325(406) & 0.609(89) & 1.122(164) & 1.16\\
    \hline
    \multirow{2}{*}{z-expansion} & $\simeq$ 0.5 & 0.702(179) &   0.973(248) & 0.99 & -2.531(415) & 0.502(19) & 1.360(52) & 0.66\\
    & $\simeq$ 1& 0.695(169) &  0.984(239) & 0.81 & -1.600(237)& 0.543(24) & 1.260(56) & 1.03\\
    \hline\hline
  \end{tabular}
\end{table*}
\end{widetext}

\begin{figure}[!ht]
     \centering
     \includegraphics[scale=0.525]{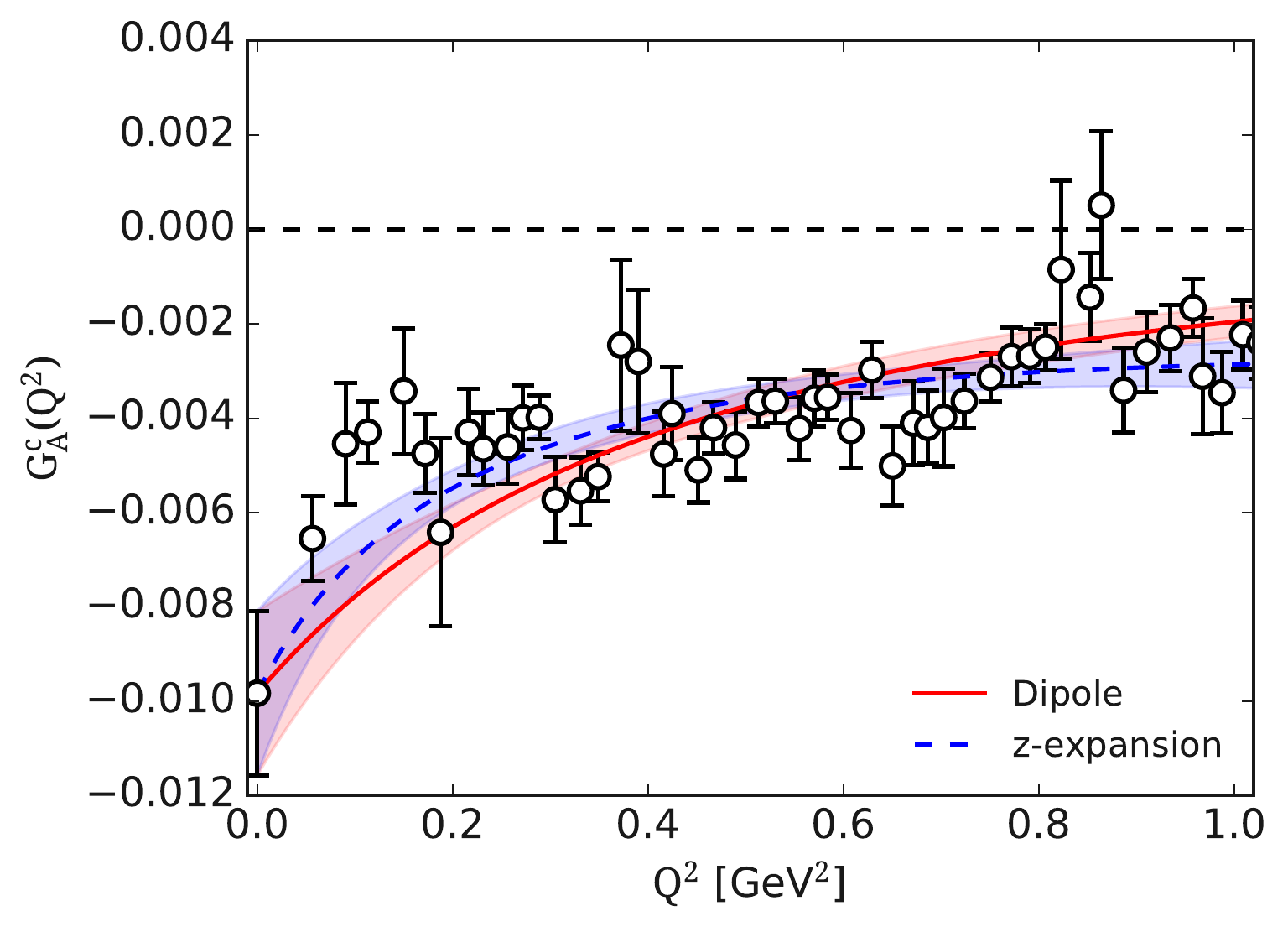}
     \includegraphics[scale=0.525]{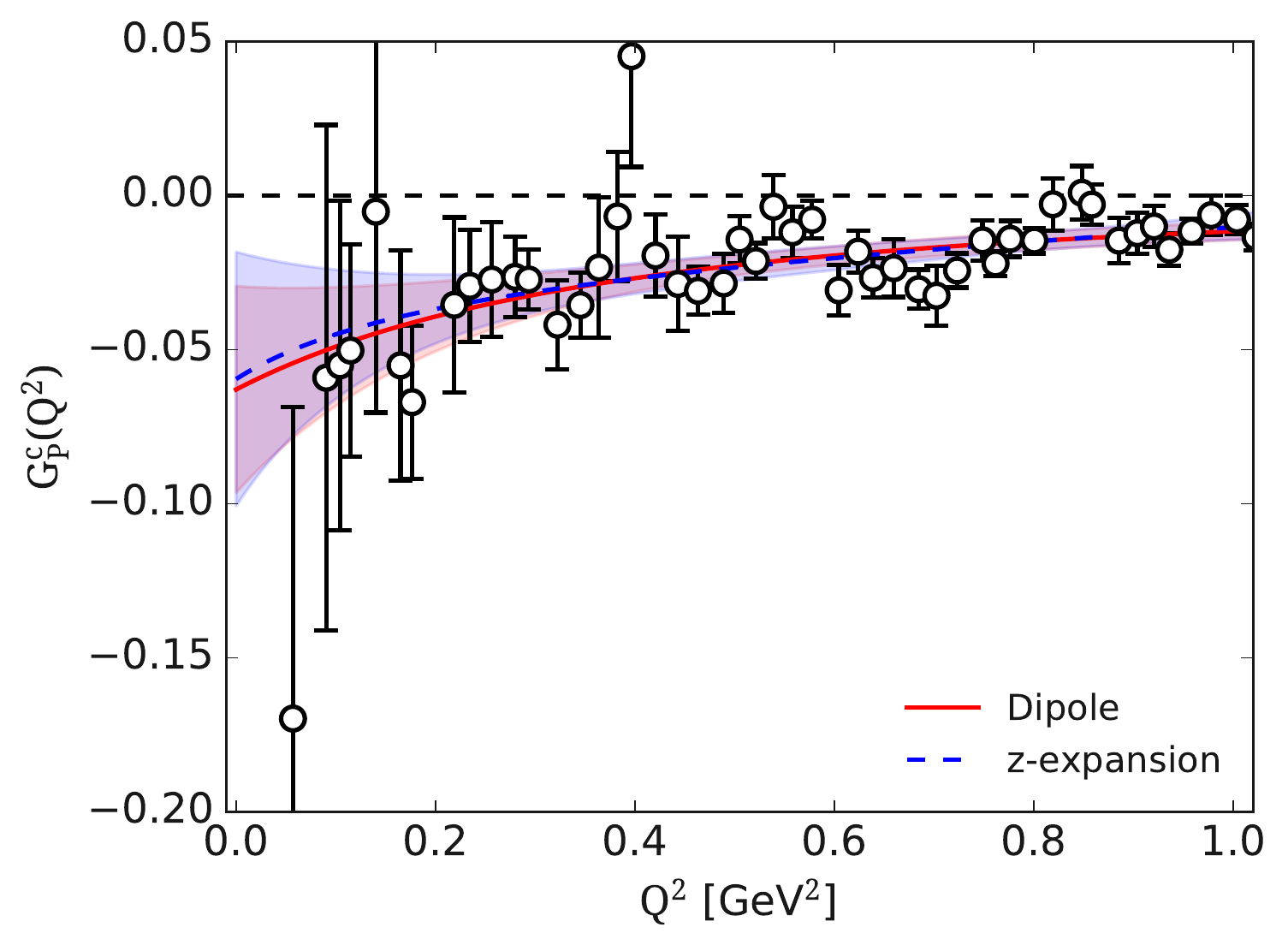}
     \caption{Results for the charm form factors, $G_A^c(Q^2)$ (left) and $G_P^c(Q^2)$ (right), as a function of $Q^2$. The notation is the same as that in Fig.~\ref{fig:GAGP_strange}.}
     \label{fig:GAGP_charm}
 \end{figure}

We follow the same analysis described for the strange form factors to extract the charm axial form factors  $G_A^c(Q^2)$ and $G_P^c(Q^2)$ that are shown in Fig.~\ref{fig:GAGP_charm}. They are both clearly negative. Performing the dipole and z-expansion fits we can determine the same parameters as in the  case of the strange form factors. The values are given  in Table~\ref{table:GA_c}. Since the slope of the z-expansion fit as $Q^2\rightarrow 0$ is steeper, the r.m.s. radius determined from the z-expansion tends to be larger as it was the case for the corresponding strange r.m.s. radius. It is worth mentioning that the z-expansion describes better the data as compared to the dipole Ansatz as indicated by the $\chi^2$/d.o.f. For the charm axial charge  we find $g_A^c=-0.0098(17)$.

\begin{widetext}
  \begin{table*}[ht!]
  \caption{Results from the charm form factors using the same notation as that in Table~\ref{table:GA_s}.}
  \label{table:GA_c}
  \vspace{0.2cm}
  \begin{tabular}{c|c|c|c|c|c||c|c|c|c}
    \hline
    Fit Type & $Q^2_{\rm max}$~[GeV$^2$] & $m_A^c$ [GeV] &  $\sqrt{\langle (r_A^c)^2 \rangle}$ [fm] & $\chi^2$/d.o.f & $G_P^c(0)$ & $m_P^c$ [GeV] &  $\sqrt{\langle (r_P^c)^2 \rangle}$ [fm] & $\chi^2$/d.o.f \\
    \hline
    \multirow{2}{*}{Dipole} & $\simeq$ 0.5 & 0.800(142) &  0.854(152) & 3.2 & -0.062(69) & 0.892(847) & 0.767(726) & 0.59\\
    & $\simeq$ 1& 0.898(132) &  0.761(112) & 2.3 & -0.063(34) & 0.867(272) & 0.788(247) & 1.02\\
    \hline
    \multirow{2}{*}{z-expansion} & $\simeq$ 0.5 & 0.534(56) & 1.280(135) &  1.6 & -0.076(40) & 0.654(127) & 1.045(203) & 0.51\\
    & $\simeq$ 1& 0.692(94) & 0.987(133) & 1.0 & -0.060(41) & 0.762(315)&0.897(369) & 0.96\\
    \hline\hline
  \end{tabular}
\end{table*}
\end{widetext}

\section{Analysis of the flavor singlet and octet axial form factors and the SU(3) symmetry breaking}\label{sec:SU3}
The determination of isoscalar and strange form factors  allows us to construct  the corresponding  SU(3) flavor octet and singlet form factors. 
 We would like to highlight that these quantities are computed for the first time  directly at the physical point.

\label{sec:fs_octet_comps}

 \begin{figure}[!ht]
     \centering
     \includegraphics[scale=0.525]{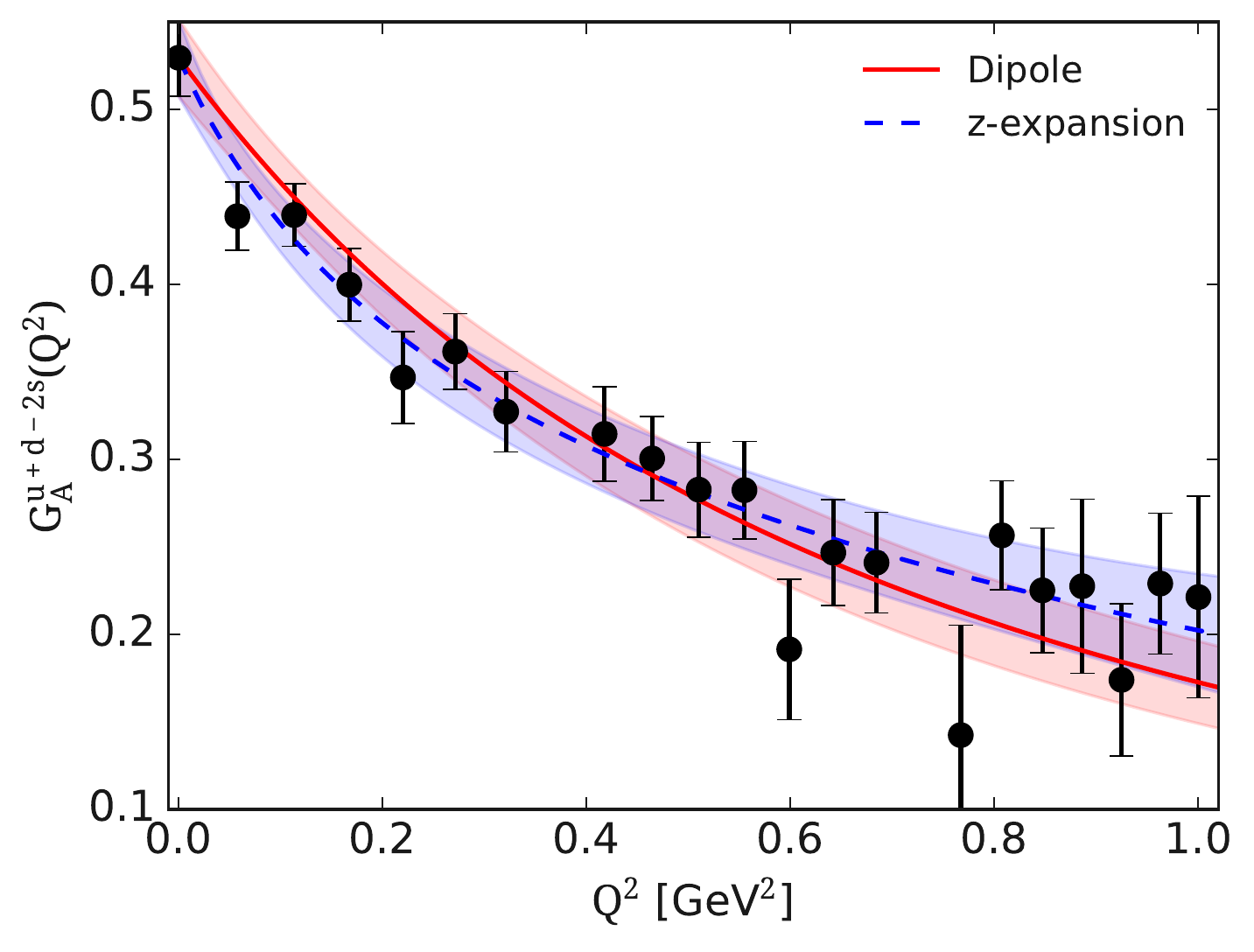}
     \includegraphics[scale=0.525]{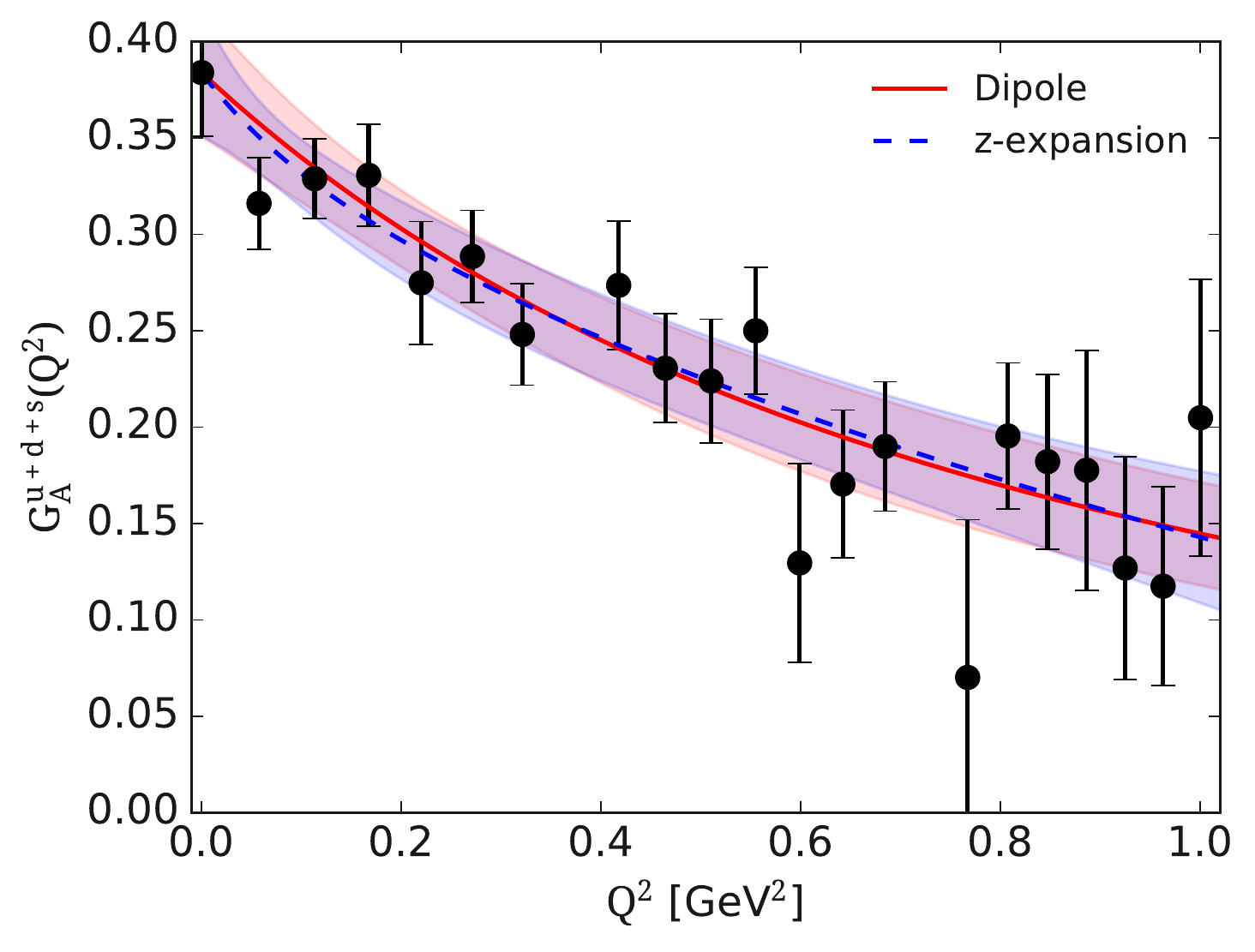}
     \caption{Results on the flavor octet (left) $G_A^{u+d-2s}(Q^2)$ and singlet (right) $G_A^{u+d+s}(Q^2)$ axial form factor  as a function of $Q^2$. }
     \label{fig:GA_octet_singlet}
 \end{figure}

\begin{figure}[!ht]
     \centering
     \includegraphics[scale=0.525]{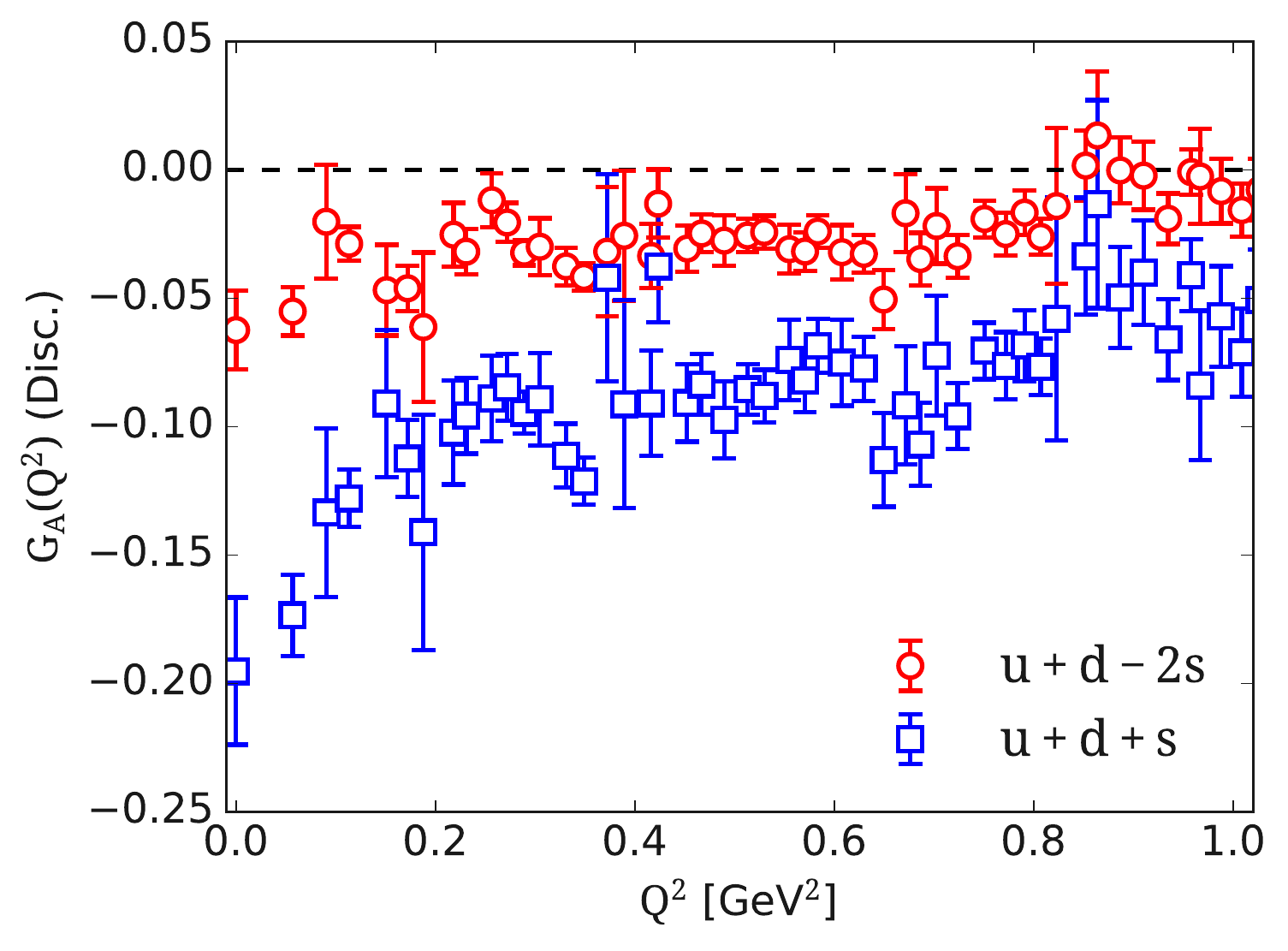}
     \includegraphics[scale=0.525]{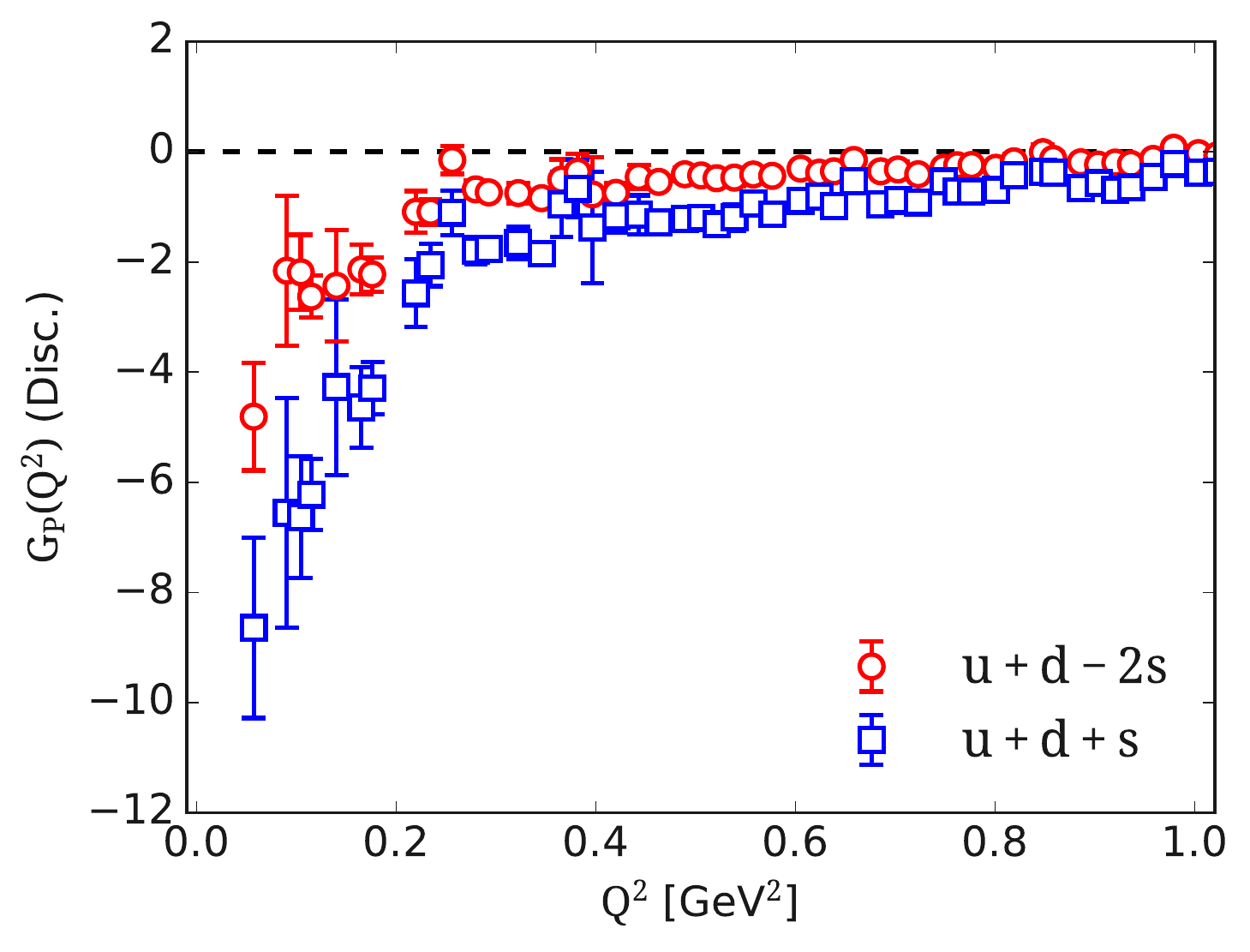}
     \caption{Results on the disconnected contribution to the SU(3) flavor octet (red circles) and singlet (blue squares) for the  axial (left) and induced pseudoscalar (right) form factors. 
     }
     \label{fig:GA_disc_oct_singlet}
 \end{figure}

In Fig.~\ref{fig:GA_octet_singlet} we present  results for the SU(3) flavor octet axial form factor $G_A^{u+d-2s}(Q^2)$  and for the singlet $G_A^{u+d+s}(Q^2)$. If SU(3) was exact, the disconnected contributions would cancel in the octet combination. In practice, we find deviations from SU(3) symmetry  especially at low $Q^2$ where the form factor is larger (see Fig.~\ref{fig:GA_disc_oct_singlet}). This demonstrates that SU(3) flavor symmetry is violated due to the different mass between light and strange quarks. This is an important result since many phenomenological analyses assume SU(3) flavor symmetry introducing an uncontrolled systematic error. We find that there  is up to 10\% breaking for the axial and up to 50\% for the induced pseudoscalar form factors. 

  Due to the suppression of  disconnected contributions in the  octet combination, as can be seen in Fig.~\ref{fig:GA_disc_oct_singlet},  $G_A^{u+d-2s}(Q^2)$ is more precise as compared to $G_A^{u+d+s}(Q^2)$ shown in Fig.~\ref{fig:GA_octet_singlet}.  The data for both octet and singlet form factors are well described by our two fit Anz\"atze, namely the dipole form and the z-expansion. The resulting values of   $\chi^2$/d.o.f are given in Table~\ref{table:GA_octet_singlet}. The value of the form factors at zero momentum transfer, gives the octet and singlet axial charges $g_A^{u+d-2s}$ and $g_A^{u+d+s}$, respectively.
We find $g_A^{u+d-2s}=0.530(22)$ and $g_A^{u+d+s}=0.384(33)$. 
These charges have been also extracted from phenomenological analyses. In Ref.~\cite{Bass:2009ed}, the authors use  polarised deep inelastic scattering data to extract $g_A^{u+d-2s}=0.46(5)$ and $g_A^{u+d+s}=0.36(3)(5)$ both in agreement with our findings but with larger uncertainties. It is worth mentioning that the analysis of Ref.~\cite{Bass:2009ed} assumes  SU(3) flavor symmetry.

 In Table~\ref{table:GA_octet_singlet}, we collect the parameters extracted from these fits.  The SU(3) flavor  octet axial mass $m_A^{u+d-2s}$  tends to have a smaller value than  the corresponding singlet, $m_A^{u+d+s}$,  which translates to a bigger octet r.m.s. radius.   However, statistical errors on the singlet quantities are large and the two values agree within the statistical errors. This is particularly true for the parameters extracted from the z-expansion where the statistical errors are even larger.
 
 \begin{widetext}
   \begin{table*}[ht!]
  \caption{The axial mass and radius determined from fitting the SU(3) flavor octet and singlet   axial form factor $G_A^{u+d-2s}(Q^2)$  and $G_A^{u+d+s}(Q^2)$, respectively, using the dipole Ansatz and the z-expansion. The notation is the same as the one in Table~\ref{table:GA_isos_extr_vals}.}
  \label{table:GA_octet_singlet}
  \vspace{0.2cm}
  \begin{tabular}{c|c||c|c|c||c|c|c}
    \hline
    Fit Type & $Q^2_{\rm max}$~[GeV$^2$] & $m_A^{u+d-2s}$ [GeV]  & $\sqrt{\langle (r_A^{u+d-2s})^2 \rangle}$ [fm] & $\chi^2$/d.o.f & $m_A^{u+d+s}$ [GeV] &  $\sqrt{\langle (r_A^{u+d+s})^2 \rangle}$ [fm] & $\chi^2$/d.o.f \\
    \hline
    \multirow{2}{*}{Dipole} & $\simeq$ 0.5 & 1.097(104)  & 0.623(59) & 1.07 & 1.255(240) &  0.545(104) & 0.68  \\
    & $\simeq$ 1& 1.154(101)   & 0.592(52) & 1.04 & 1.261(188)  &  0.542(81) & 0.65 \\
    \hline
    \multirow{2}{*}{z-expansion} & $\simeq$ 0.5 & 0.876(121)  & 0.780(108) & 0.45 & 1.016(335)  &  0.673(221) & 0.50   \\
    & $\simeq$ 1& 0.898(134)    & 0.761(113) &  0.57 & 1.051(359) &  0.650(221) & 0.59\\
    \hline\hline
  \end{tabular}
\end{table*}
 \end{widetext}

\begin{figure}[!h]
     \centering
     \includegraphics[scale=0.525]{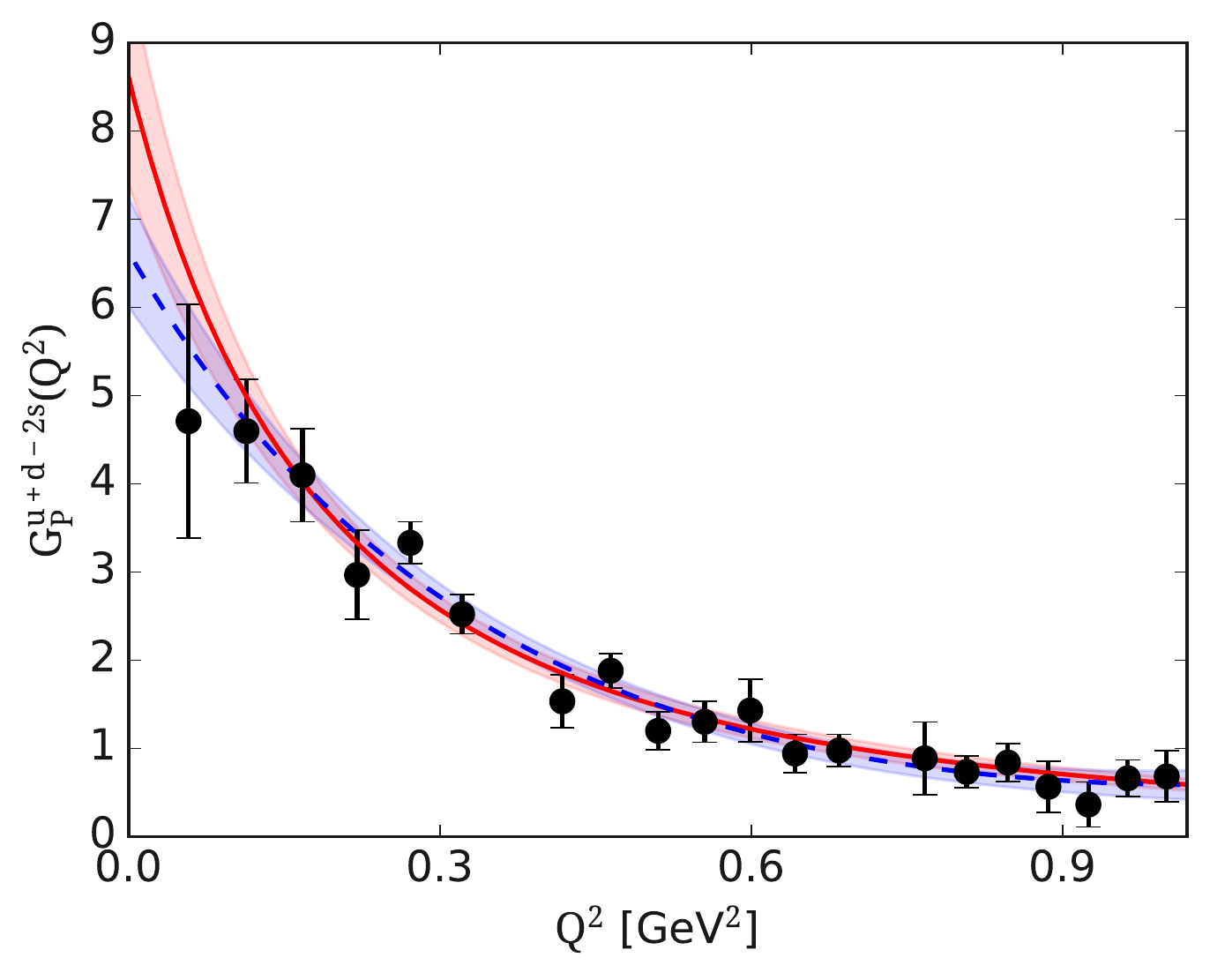}
     \includegraphics[scale=0.525]{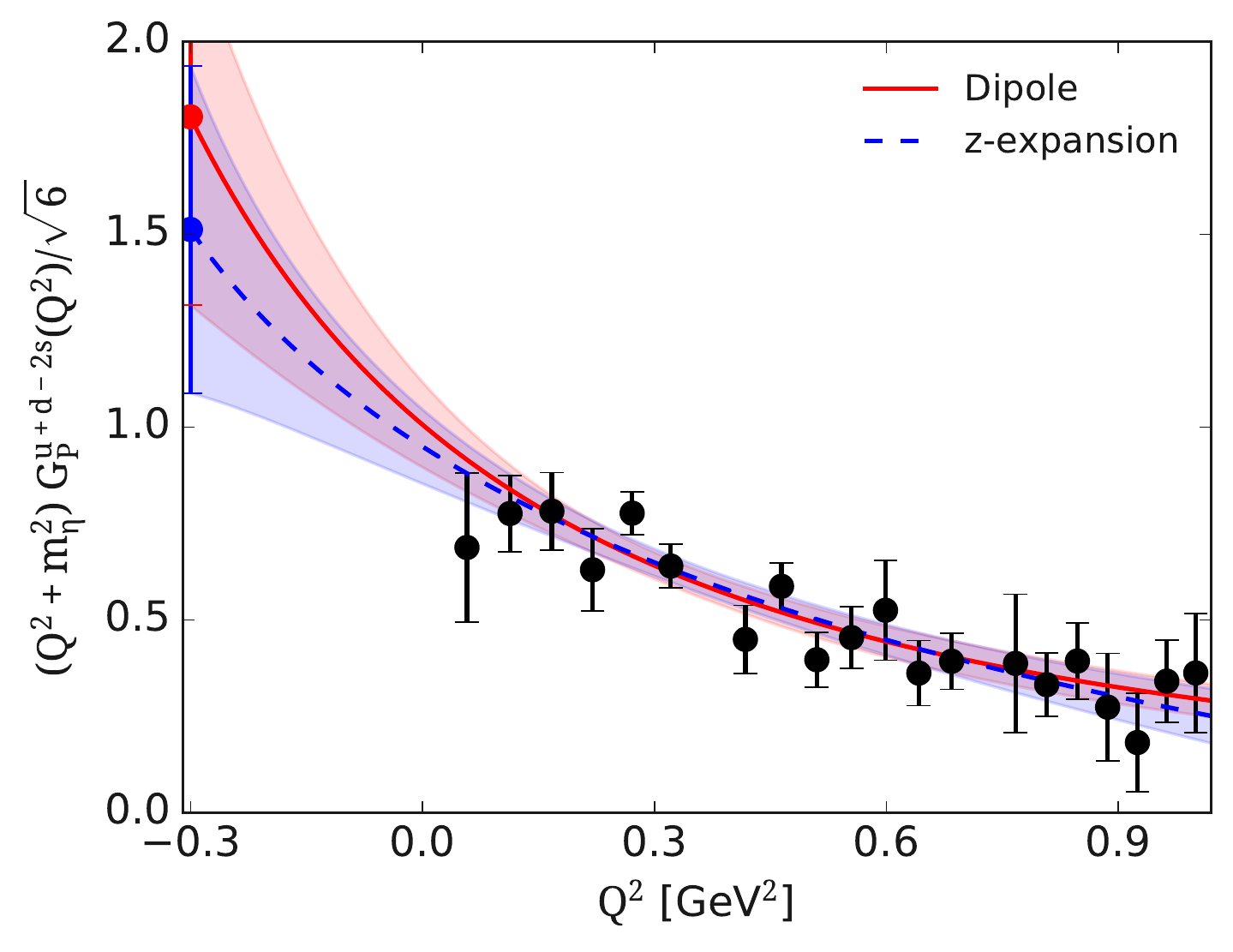}
     \caption{\emph{Left:} Results on the induced flavor octet pseudoscalar form factor. The red and blue bands, which are overlapping,  show the results of the  fits to the dipole form and using the  z-expansion. \emph{Right:} Results on the induced pseudoscalar form factor shown in the left panel after canceling  the $\eta$-meson pole,  i.e. we show $(m_\eta^2+Q^2)G_P^{u+d-2s}(Q^2)$ as a function of $Q^2$. }
     \label{fig:GP_octet}
 \end{figure}

The $Q^2$-dependence of the induced octet pseudoscalar form factors $G_P^{u+d-2s}(Q^2)$  is shown in Fig.~\ref{fig:GP_octet} with the corresponding extracted parameters provided in Table~\ref{table:GP_octet}. It is well-known that the isovector induced 
pseudoscalar form factor, $G_P^{u-d}(Q^2)$ has a pion pole behavior. Results on this form factor  using the same ensemble were reported in Ref.~\cite{Alexandrou:2020okk}. By similar arguments, the SU(3) flavor octet form factor  $G_P^{u+d-2s}(Q^2)$ is expected to have an $\eta$ pole behaviour. Since the $\eta$-meson has a much larger mass compared to the mass of the pion, the  relations that hold  in the chiral limit for $G_P^{u-d}(Q^2)$ are expected to be significantly violated in this case.

 \begin{widetext}
   \begin{table*}[h!]
  \caption{Parameters extracted from the induced pseudoscalar form factor for the flavor octet combination.}
  \label{table:GP_octet}
  \vspace{0.2cm}
  \begin{tabular}{c|c||c|c|c|c|}
    \hline
    Fit Type & $Q^2_{\rm max}$~[GeV$^2$] & $G_P^{u+d-2s}(0)$ & $m_P^{u+d-2s}$ [GeV]  & $\sqrt{\langle (r_P^{u+d-2s})^2 \rangle}$ [fm] & $\chi^2$/d.o.f  \\
    \hline
    \multirow{2}{*}{Dipole} & $\simeq$ 0.5 &  7.194(1.072) & 0.691(67) & 0.989(95) & 1.2 \\
    & $\simeq$ 1&  8.587(1.204)  & 0.602(47) & 1.135(88) &   0.85 \\
    \hline
    \multirow{2}{*}{z-expansion} & $\simeq$ 0.5 & 6.024(947)  & 0.537(91) & 1.273(216) & 0.77   \\
    & $\simeq$ 1& 6.621(618)  & 0.484(20) & 1.411(48)  & 0.59 \\
    \hline\hline
  \end{tabular}
\end{table*}
 \end{widetext}

In Fig.~\ref{fig:GP_octet} we show also results on $(m_\eta^2+Q^2)G_P^{u+d-2s}(Q^2)$ that cancel the $\eta$-meson pole, as well as the dipole and z-expansion fits. This allows the extraction of the eta-nucleon coupling $g_{\eta NN}$ in analogy to the determination of $g_{\pi NN}$ since
\begin{equation}
  g_{\eta NN} =  \frac{\displaystyle \lim_{Q^2 \rightarrow -m_\eta^2} (Q^2 + m_\eta^2) G_P^{u+d-2s}}{ 4 m_N F_\eta^8},
    \label{Eq:getaNN}
\end{equation}
where  $F_\eta^8$ is the decay constant of the $\eta$ meson and $m_\eta$ its mass. We note that the mixing with the $\eta^\prime$ has been neglected in Eq.~(\ref{Eq:getaNN}).
The $\eta$ decay constant $F_\eta^8 $ can be determined directly in lattice QCD in an analogous manner to the computation of $F_\pi$~\cite{Ottnad:2017bjt}. This will be computed for the current ensemble in a future work. 
Here,  we use the value of $F_\eta^8 $  determined from phenomenology~\cite{Feldmann:1999uf} to extract the coupling constant 
\begin{eqnarray}
    g_{\eta NN} &=& 4.5(1.2) \qquad(\text{dipole})\\
    g_{\eta NN} &=& 3.7(1.0)\qquad(\text{z-expansion}).
\end{eqnarray}
The extrapolation to $Q^2=-m_\eta^2 $ is shown in the right panel of Fig.~\ref{fig:GP_octet}. As can be seen, the fact that one needs  to perform a large  extrapolation in the negative $Q^2$ region increases the statistical uncertainty as compared to the isovector case. 
The values extracted are in agreement with the ones extracted from phenomenological studies~\cite{Nasrallah:2005hn,Feldmann:1999uf,Dumbrajs:1983jd}.

\begin{figure}[!ht]
    \centering
    \includegraphics[scale=0.525]{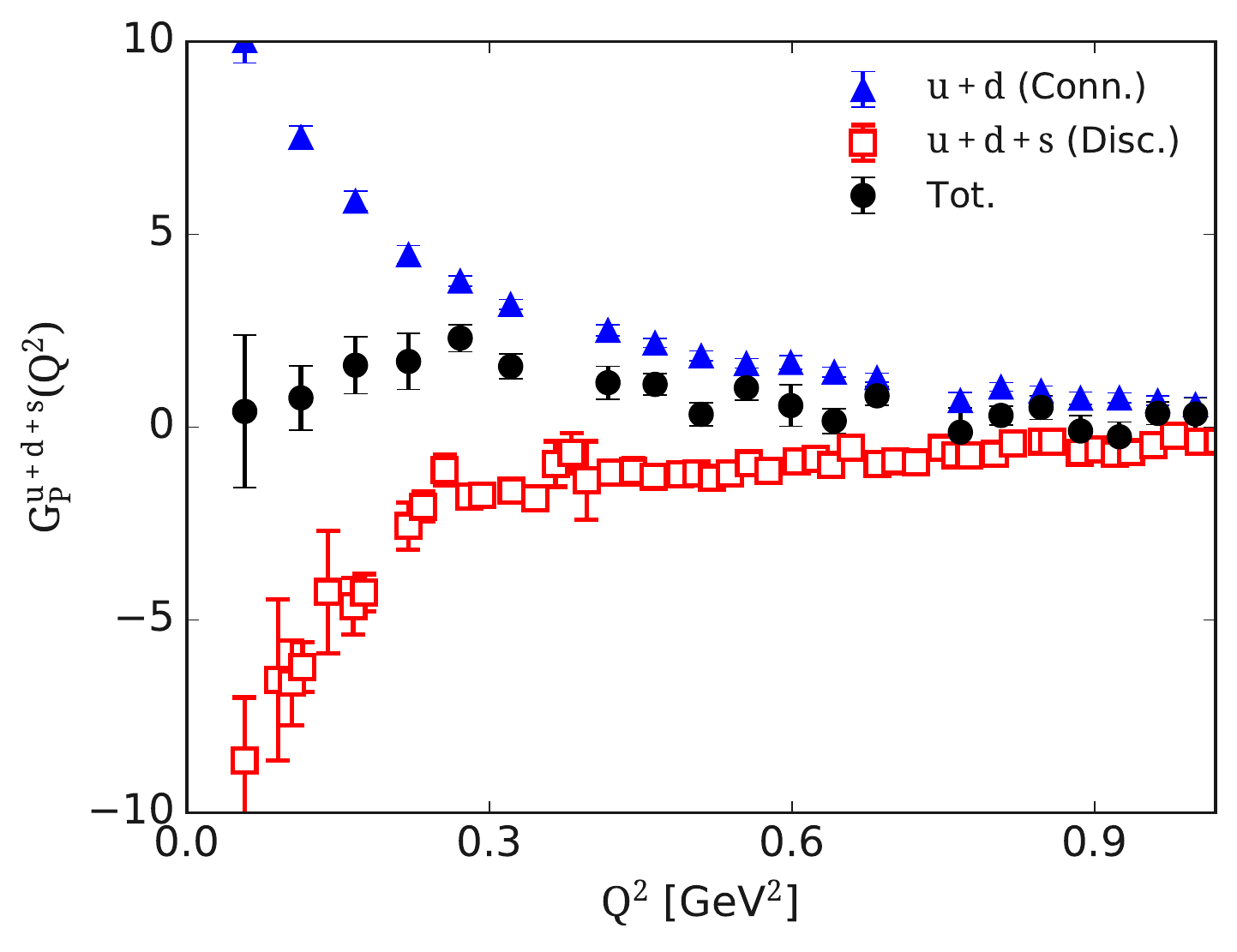}
    \caption{Results on the flavor singlet induced pseudoscalar form factor. The notation is the same as that  in Fig.~\ref{fig:GP_isos_f1}.
    }
    \label{fig:GP_singlet}
\end{figure}

If one defines a Goldberger-Treiman discrepancy for the octet in a similar manner as done for the isovector combination
\begin{equation}
    \Delta_{GT}^{8} = 1 - \frac{g_A^{u+d-2s} m_N}{g_{\eta NN} F_\eta^8},
\end{equation}
can assess how much the Goldberger-Treiman relation is violated in this case. We find that
\begin{eqnarray}
    \Delta_{GT}^{8} &=& 0.42(12) \qquad(\text{dipole})\\
     \Delta_{GT}^{8} &=& 0.50(14) \qquad(\text{z-expansion}).
\end{eqnarray}
We find a violation of about 40-50\% for the octet combination of  $\Delta_{GT}$, which is much larger than the 2\% determined for the isovector combination~\cite{Alexandrou:2020okk}. This is a consequence of the large $\eta$-meson mass.  The flavor singlet  induced pseudoscalar form factor $G_P^{u+d+s} (Q^2)$ is  noisy because of the disconnected contributions are large and of opposite sign to the connected partly canceling each other, as can be seen in Fig.\ref{fig:GP_singlet}.

\section{The up and down  axial and induced pseudoscalar form factors} \label{sec:FL_decomp}
Having determined  the isovector~\cite{Alexandrou:2020okk} and isoscalar form factors we can disentangle the up and down quark contributions to the these form factors.
\begin{figure}[!h]
     \centering
     \includegraphics[scale=0.525]{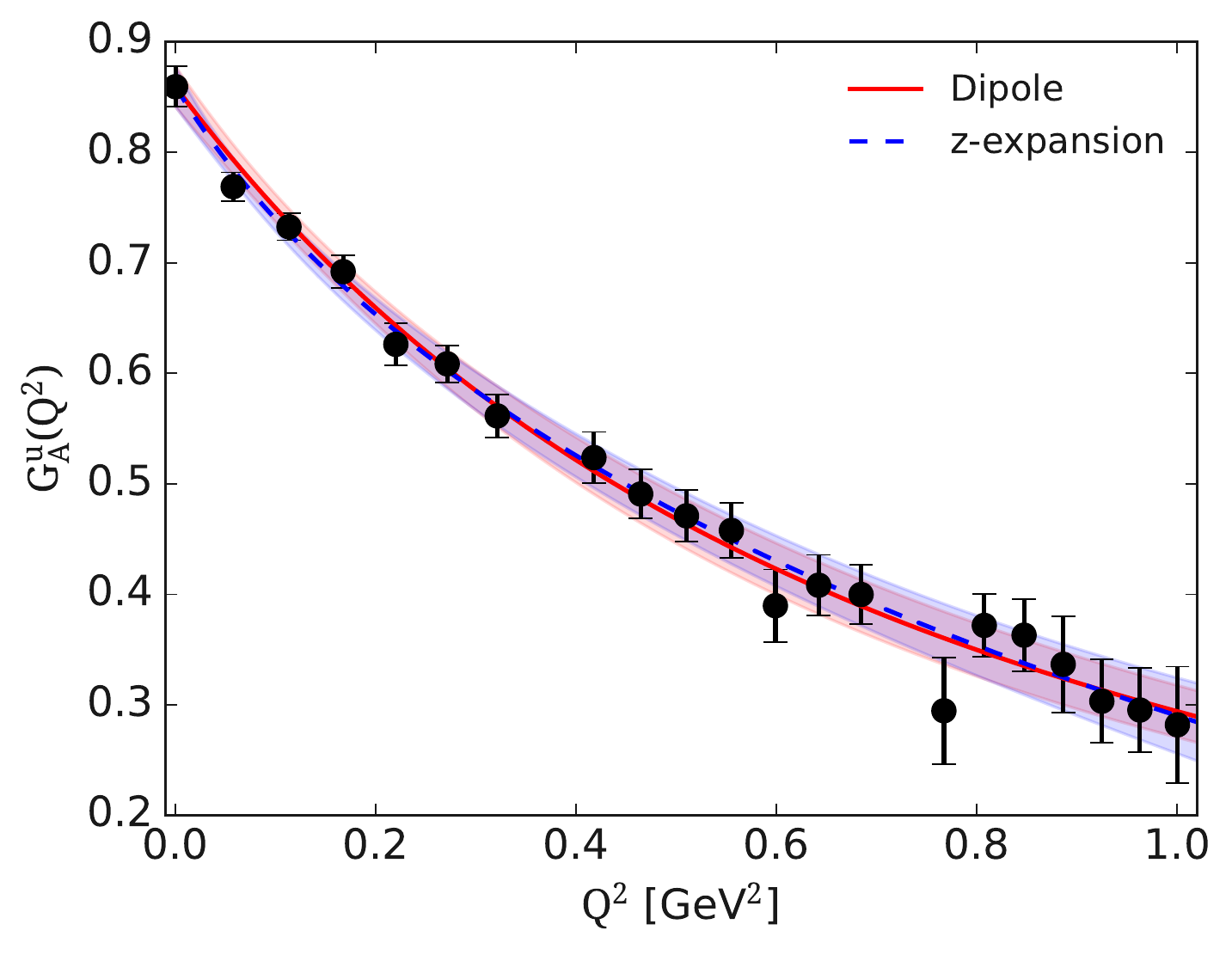}
          \includegraphics[scale=0.525]{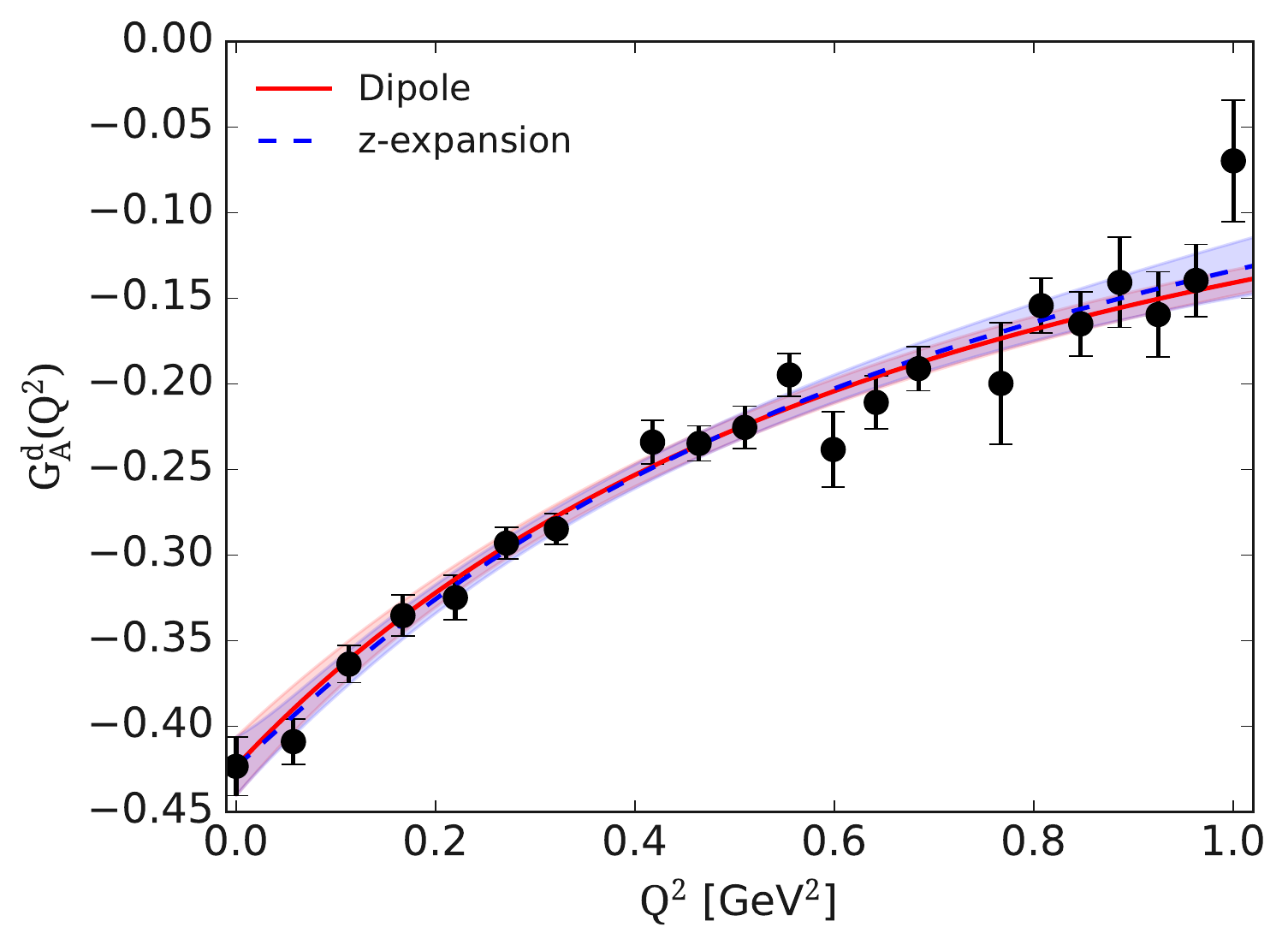}
     \caption{Results for  the up  (left) and down (right)  quark axial form factors $G_A^u(Q^2)$ and $G_A^d(Q^2)$  an as a function of $Q^2$. The red band shows the results from the dipole fit and the blue band using the z-expansion.}
     \label{fig:GA_up_down}
 \end{figure}

   In Fig.~\ref{fig:GA_up_down} we show results for the  up and down quark axial form factors, $G_A^u(Q^2)$  and $G_A^d(Q^2)$ as a function of $Q^2$. $G_A^u(Q^2)$ is found to be positive, while  $G_A^d(Q^2)$ is negative and about half in magnitude. The axial up and down quark charges obtained at $Q^2=0$ are  $g_A^u=0.859(18)$ and $g_A^d=-0.423(17)$ in agreement with the values found in Ref.~\cite{Alexandrou:2019brg}. Since the value of the form factors at zero momentum is known, we use it to eliminate one fit parameter in the jackknife analysis.
 The  values of the up and down quark axial masses and r.m.s. radii extracted from the dipole fit and using the z-expansion are given in Table~\ref{table:GA_up_down}. We find that $m_A^u\sim m_A^d$ and $\sqrt{\langle (r_A^u)^2 \rangle} \sim \sqrt{\langle (r_A^d)^2 \rangle}$ within statistical errors. 

 \begin{widetext}
   \begin{table*}[h!]
  \caption{The axial mass and radius determined from fitting  $G_A^{u}(Q^2)$  and $G_A^{d}(Q^2)$, using the dipole Ansatz and the z-expansion. The notation is the same as that in Table~\ref{table:GA_octet_singlet}.}
  \label{table:GA_up_down}
  \vspace{0.2cm}
  \begin{tabular}{c|c|c|c|c||c|c|c}
    \hline
    Fit Type & $Q^2_{\rm max}$~[GeV$^2$] & $m_A^u$ [GeV]  & $\sqrt{\langle (r_A^u)^2 \rangle}$ [fm] & $\chi^2$/d.o.f & $m_A^d$ [GeV]  & $\sqrt{\langle (r_A^d)^2 \rangle}$ [fm] & $\chi^2$/d.o.f\\
    \hline

    \multirow{2}{*}{Dipole} & $\simeq$ 0.5 & 1.179(70) &   0.580(34) & 0.62 & 1.174(65) &  0.582(32) & 0.58 \\
                            & $\simeq$ 1   & 1.187(65)  &   0.576(32) & 0.52 & 1.168(54)  & 0.585(27) & 0.81 \\
    \hline
    \multirow{2}{*}{z-expansion} & $\simeq$ 0.5 & 1.050(118) &  0.651(73) & 0.37 & 1.336(341)  &   0.512(130) & 0.39  \\
                                 & $\simeq$ 1&  1.069(122) &  0.639(73) & 0.41 & 1.312(329) &   0.521(131) & 0.72 \\
    \hline\hline
  \end{tabular}
\end{table*}
\end{widetext}

  \begin{figure}[!h]
     \centering
     \includegraphics[scale=0.525]{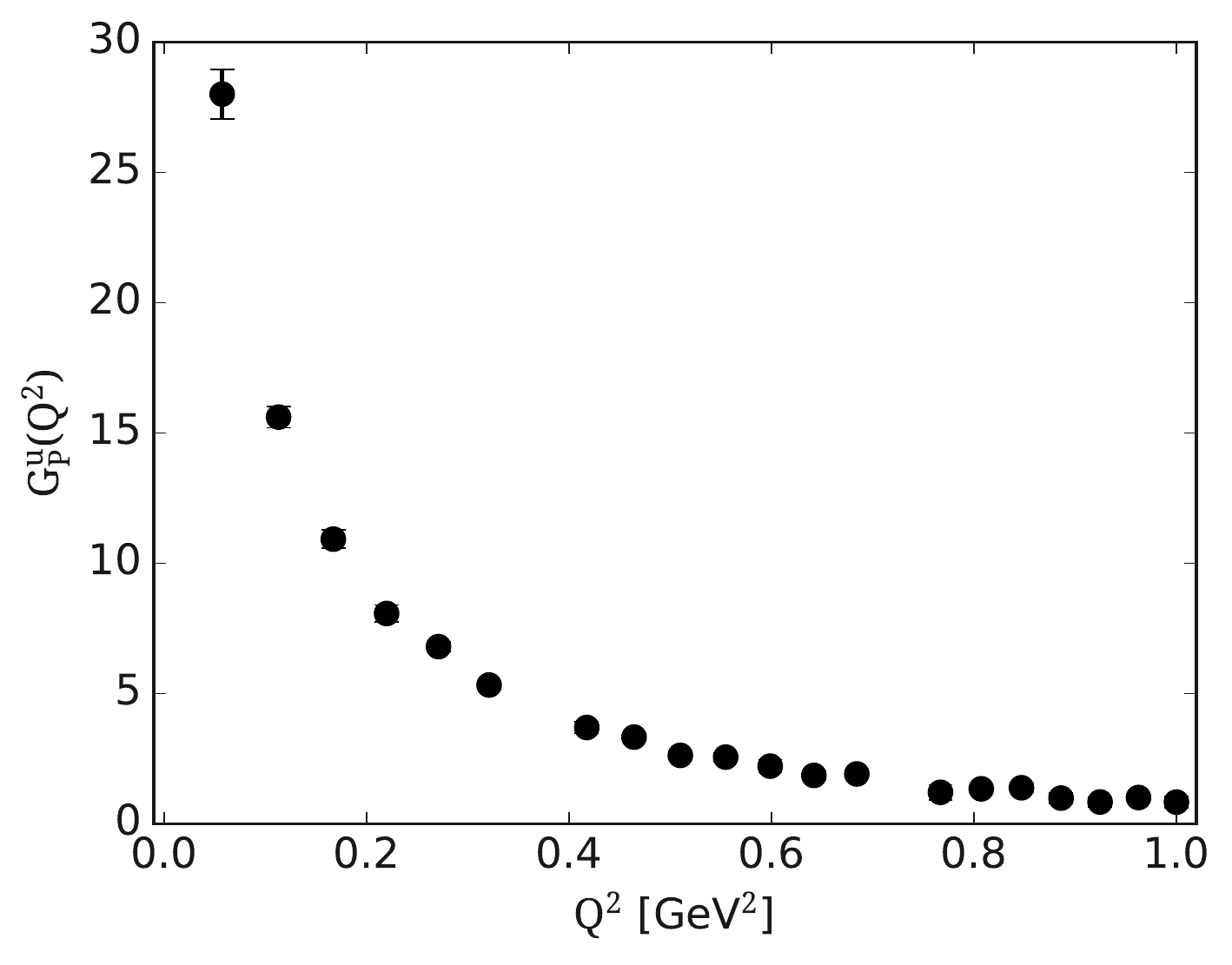}
     \includegraphics[scale=0.525]{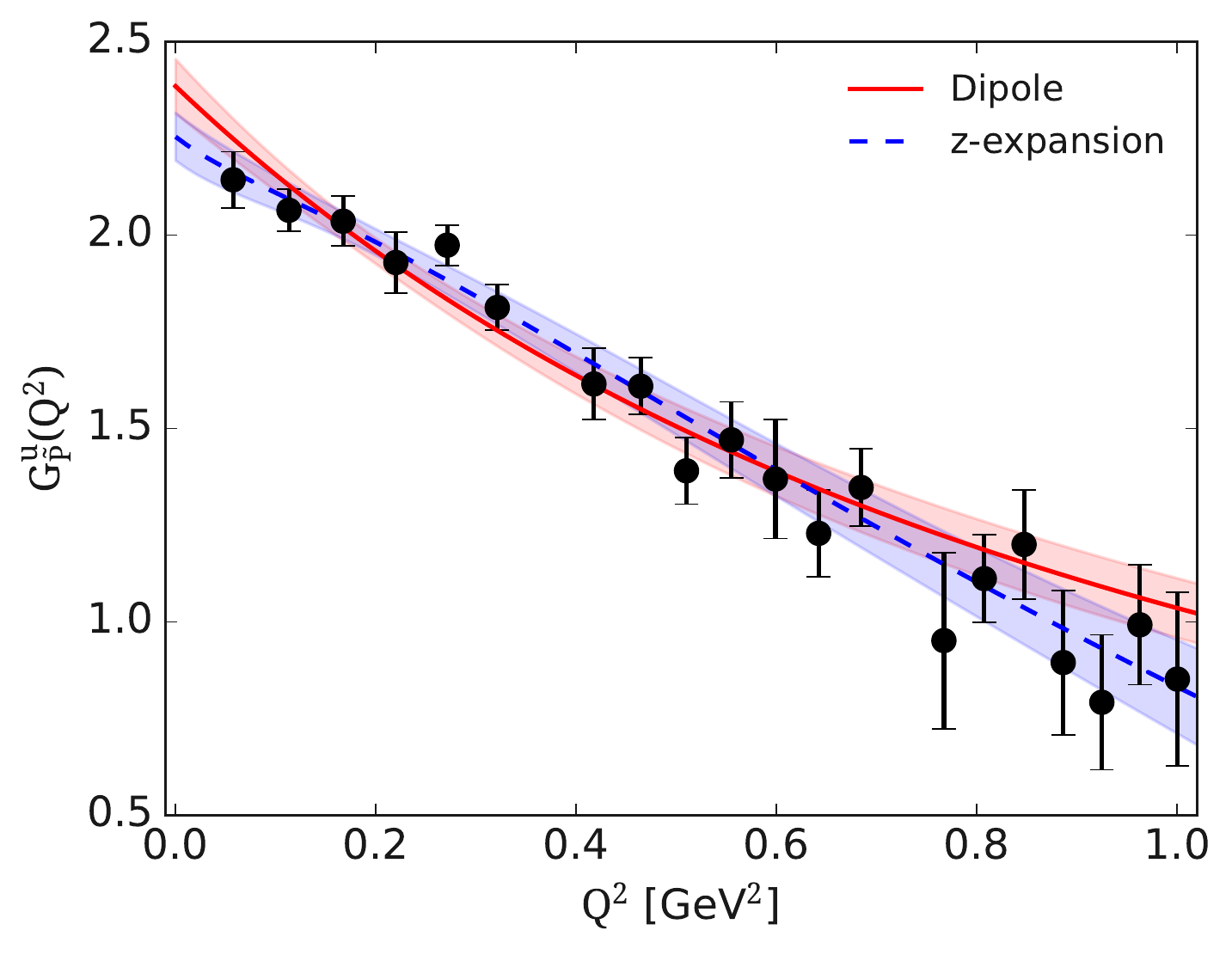}
     \caption{\emph{Left:} Results for  the  up quark induced pseudoscalar form factor $G_P^u(Q^2)$  as a function of $Q^2$.
     \emph{Right:} Results for the up quark induced pseudoscalar form factor after cancelling the pion pole using Eq.~\eqref{Eq:Gtilde}.
     The notation is the same as Fig.~\ref{fig:GA_up_down}.}
     \label{fig:GP_up}
  \end{figure}
 \begin{figure}[!h]
     \centering
     \includegraphics[scale=0.525]{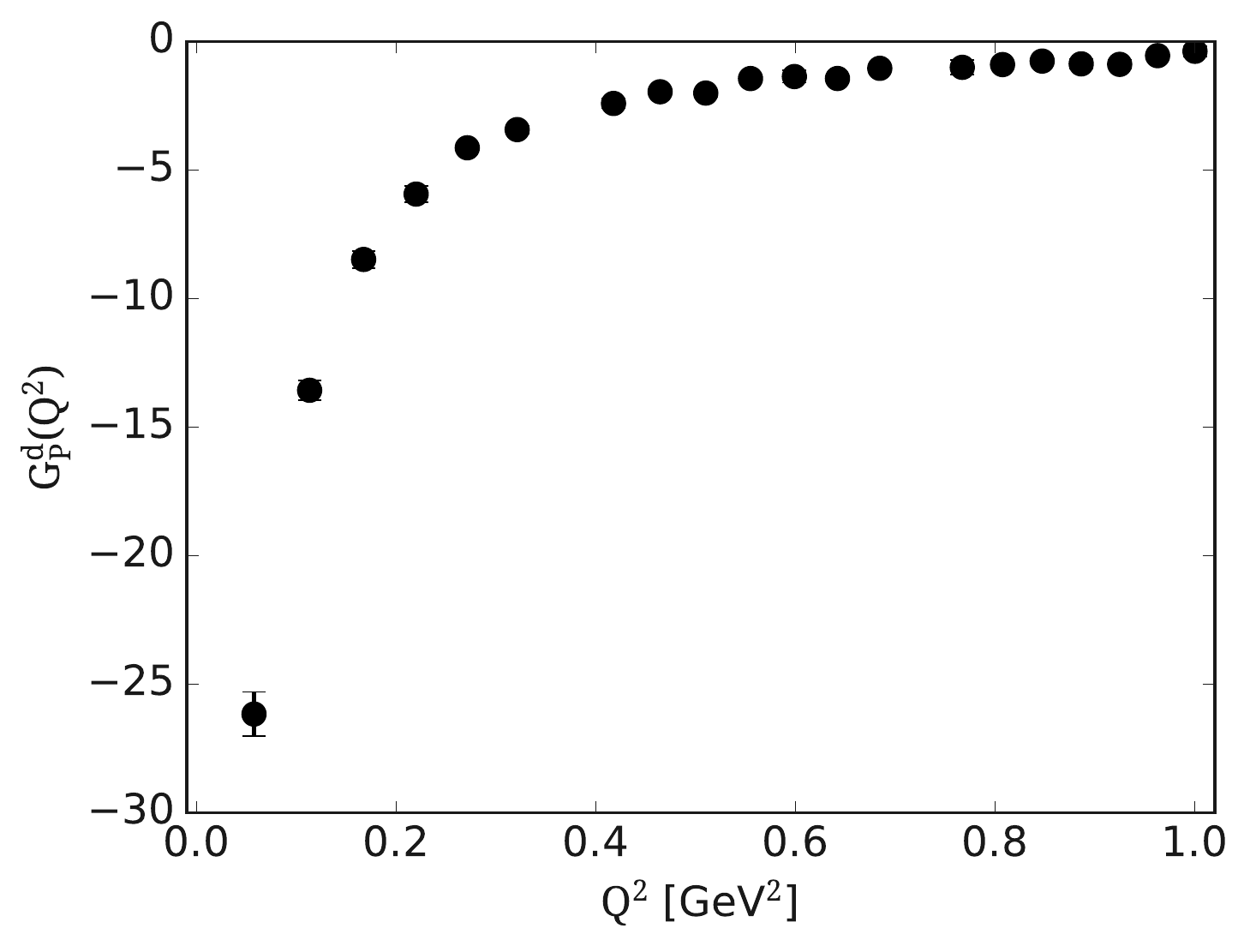}
     \includegraphics[scale=0.525]{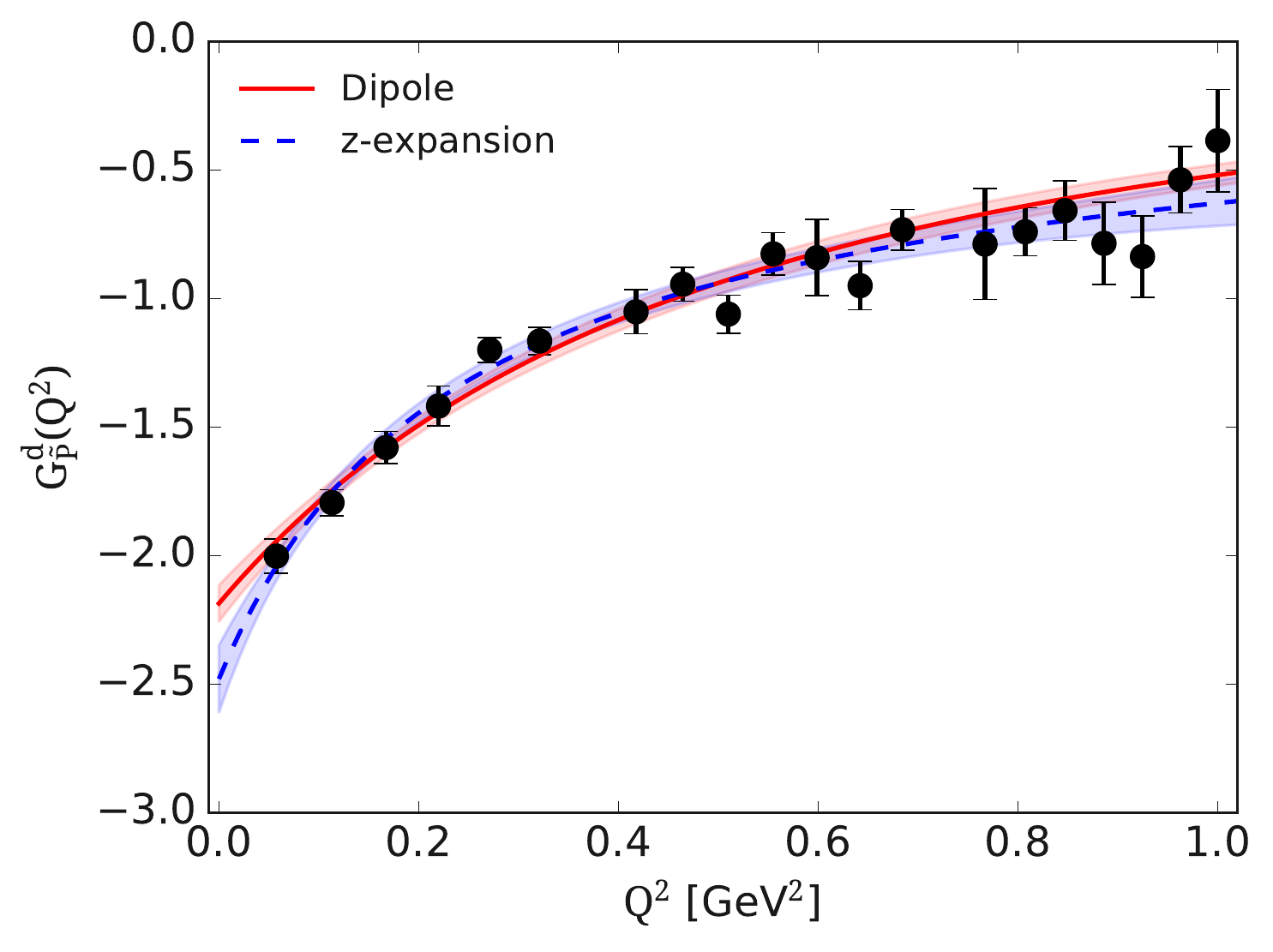}
     \caption{Results for  the down quark induced pseudoscalar form factor as a function of $Q^2$.  The notation is the same as Fig.~\ref{fig:GP_up}.}
     \label{fig:GP_down}
 \end{figure}
 In Figs.~\ref{fig:GP_up} and \ref{fig:GP_down} we show results on the up and down quark induced pseudoscalar form factors. The large slope
 observed for these form factors as $Q^2 \rightarrow 0$ is due to the presence of the pion pole. Before we  fit them  with a dipole form and the z-expansion we eliminate the pion pole  and consider instead  
 \begin{equation}
 G_{\tilde{P}}(Q^2) \equiv (Q^2+m_\pi^2) \; G_P(Q^2)
 \label{Eq:Gtilde}
 \end{equation}
  for the fits. Note that $G_{\tilde{P}}(Q^2)$ has units of GeV$^2$.  Like in   the case of $G_A^{u,d}(Q^2)$, $G_P^{u}(Q^2)$ is positive and
 $G_P^{d}(Q^2)$ is negative. However, unlike $G_A^{u,d}(Q^2)$, both $G_P^{u}(Q^2)$ and $G_P^{d}(Q^2)$ have  similar magnitude. 
 For the case of $G_{\tilde{P}}(Q^2)$, the $Q^2$-dependence is different being more linear for the up quark as compared to down quark.  The dipole fits to both $G_{\tilde{P}}^u(Q^2$ and  $G_{\tilde{P}}^{d}(Q^2)$ do not describe the curvature as well  as the z-expansion fit does,  producing more curvature for the former and less for the latter as compared to the lattice QCD  data.

In Table~\ref{table:GP_up_down}, we provide the parameters extracted from the up and down induced pseudoscalar form factors. To relate the parameters, we utilize the relations
\begin{equation}
G_P(0) = \frac{G_{\tilde{P}}(0) }{ m_\pi^2},\,\, {\rm and}\,\,
    \langle r^2_P \rangle = \frac{6}{m_\pi^2} + \langle r^2_{\tilde{P}} \rangle.
    \label{Eq:tilde2}
\end{equation}

 \begin{widetext}
    \begin{table*}[ht!]
  \caption{Extracted radii and dipole masses for the up and down quark induced pseudoscalar form factors $G_P^{u,d}(Q^2)$. Note that we  use Eqs.~\eqref{Eq:Gtilde} and \eqref{Eq:tilde2} to relate the parameters extracted from $G_{\tilde{P}}(Q^2)$ of Figs.~\ref{fig:GP_up},~\ref{fig:GP_down} to those of $G_P(Q^2)$. }
  \label{table:GP_up_down}
  \vspace{0.2cm}

  \begin{tabular}{c|c|c|c|c|c||c|c|c|c|}
    \hline
    Fit Type & $Q^2_{\rm max}$~[GeV$^2$] & $G_{P}^u$(0) & $m_{P}^u$ [GeV]  & $\sqrt{\langle (r_{P}^u)^2 \rangle}$ [fm] & $\chi^2$/d.o.f & $G_{{P}}^d$(0) & $m_{{P}}^d$ [GeV]  & $\sqrt{\langle (r_{{P}}^d)^2 \rangle}$ [fm] & $\chi^2$/d.o.f\\
    \hline

    \multirow{2}{*}{Dipole} & $\simeq$ 0.5 & 119(4)  &  0.194(1)  & 3.526(18) & 0.90 & -122(4) & 0.191(1) & 3.587(19) & 0.62\\
                            & $\simeq$ 1   & 125(4)  &  0.193(1)  & 3.536(18) & 1.24 & -115(4) & 0.191(1) & 3.571(19) & 1.30\\
    \hline
    \multirow{2}{*}{z-expansion} & $\simeq$ 0.5 & 119(3)  & 0.195(1)  & 3.503(18) & 0.60  &  -126(7)  & 0.195(1) & 3.504(18) & 0.38 \\
                                 & $\simeq$ 1   &  119(3) & 0.195(1)  & 3.503(18) & 0.63  &   -130(7) & 0.191(1) & 3.504(18) & 0.72\\
    \hline\hline
  \end{tabular}

\end{table*}
\end{widetext}

\section{Final results}\label{Sec:FinalRes}
In this section we collect our  final results extracted from the fits to 
the axial and induced pseudoscalar form factors for the various flavor combinations. Results are provided using the 
z-expansion given in Eq.~\eqref{Eq:zExp}, since in most cases it fits better the form factors as in, e.g., $G_{\tilde P}^{u,d}(Q^2)$.  Fits to the dipole Ansatz  are used as a determination of the systematic error due to the choice of the fit form, by taking the  difference between the z-expansion and dipole fit values. In addition, we use the two different $Q^2$  fit ranges, namely $Q^2\simeq0.5$ and $Q^2\simeq1$~GeV$^2$ to extract a systematic due to the fit range dependence. We quote as the parameters extracted using as upper range  $Q^2\simeq1$~GeV$^2$ in the fit and the difference between 
the mean values extracted using the two ranges as the systematic error. 

 \begin{widetext}
\begin{table*}[h!]
    \caption{Final results of this work. In the first column, we give   the quark flavor combination considered, in the  second and third columns  the axial mass and r.m.s radii and in the rest three columns the value of the form factor extrapolated to $Q^2=0$, the dipole mass and r.m.s radii extracted from fitting the   induced pseudoscalar form factor. The first error is purely statistical, the second is a systematic due to different fit ranges and the third is a  systematic due to the two different forms used to fit the $Q^2$-dependence.}
    \centering
    \begin{tabular}{c||c|c|c|c|c}
        Comb. & $m_A$ [GeV]& $\sqrt{\langle r_A^2 \rangle}$ [fm] & $G_P(0)$ & $m_P$ [GeV]& $\sqrt{\langle r_P^2 \rangle}$ [fm]\\
        \hline
         u & \,\,1.069(122)(19)(118)\,\, &\,\, 0.639(73)(12)(63) \,\,&\,\, 119(3)(0)(6)  \,\,&\,\, 0.195(1)(0)(2) \,\,&\,\, 3.503(18)(0)(33)\\
         d & 1.312(329)(24)(144) & 0.521(131)(9)(64) & -130(7)(4)(15)& 0.195(1)(4)(0) & 3.504(18)(0)(67)\\
         s & 0.695(169)(7)(297)  & 0.984(239)(12)(295) & -1.600(237)(931)(275) & 0.543(24)(41)(66) & 1.260(56)(100)(138) \\
         c & 0.692(94)(158)(206) & 0.987(133)(293)(226) & -0.060(41)(16)(3) & 0.762(315)(108)(105) & 0.897(369)(148)(109)  \\
       u+d & 0.975(234)(26)(241) & 0.701(168)(19)(139) & -11.0(7.6)(4.0)(16.0) & 0.94(46)(24)(97) & 1.37(67)(35)(1.40)\\
    u+d-2s & 0.898(134)(22)(256) & 0.761(113)(19)(169) & 6.621(618)(597)(1.966) & 0.484(20)(53)(118) & 1.411(48)(138)(276)\\
     u+d+s & 1.051(359)(35)(210) & 0.650(221)(23)(108) & -12.6(7.6)(4.1)(16.0) & 0.91(40)(21)(85) & 1.33(59)(31)(1.20)
    \end{tabular}
    \label{table:FinalRes1}
\end{table*}
\end{widetext}

In Table~\ref{table:FinalRes1}, results for the axial and induced pseudoscalar masses and r.m.s radii for the  quark flavor combinations considered are provided. It is worth mentioning that this is the first time that these radii are determined for each quark flavor separately but also for the octet and singlet combinations providing us with detailed information on the structure properties of the nucleon. A notable finding is that the axial  strange and charm r.m.s radii tend to be larger than those for the light quarks. However, the  uncertainties are still large and we would need to improve the accuracy in order to draw any definite conclusion.
The values of $G_P^{u,d}(0)$ for both up and down quarks are very large compared to the rest due to the presence of the pion pole. This sharp rise of these
two form factors is reflected in the extracted r.m.s radii which are 
significantly larger than all the rest. 

From the flavor octet combination we can determine the pseudoscalar $\eta$-meson-nucleon coupling. The value is given in Eq.~\eqref{FinalRes2}, where the first error is  statistical and the second is a systematic due to the fit form used.  This  is the first determination at the physical point. It is, however, in agreement with a previous lattice QCD study~\cite{Green:2017keo} for an ensemble with pion mass of 317~MeV. The Goldberger-Treiman discrepancy $\Delta_{GT}$ is also determined for the octet combination and it is found to be 50\% which highlights that such relations are badly broken for mesons with much larger mass than the pion.

    \begin{equation}
        g_{\eta NN}= 3.7(1.0)(0.7),\,\,\,\Delta_{GT}^{u+d-2s}=0.50(14)(8) 
  \label{FinalRes2}
    \end{equation}

\section{Comparison with previous studies}\label{sec:Comp}

\begin{widetext}
 \begin{figure}[h!]
     \centering
     \includegraphics[scale=0.4]{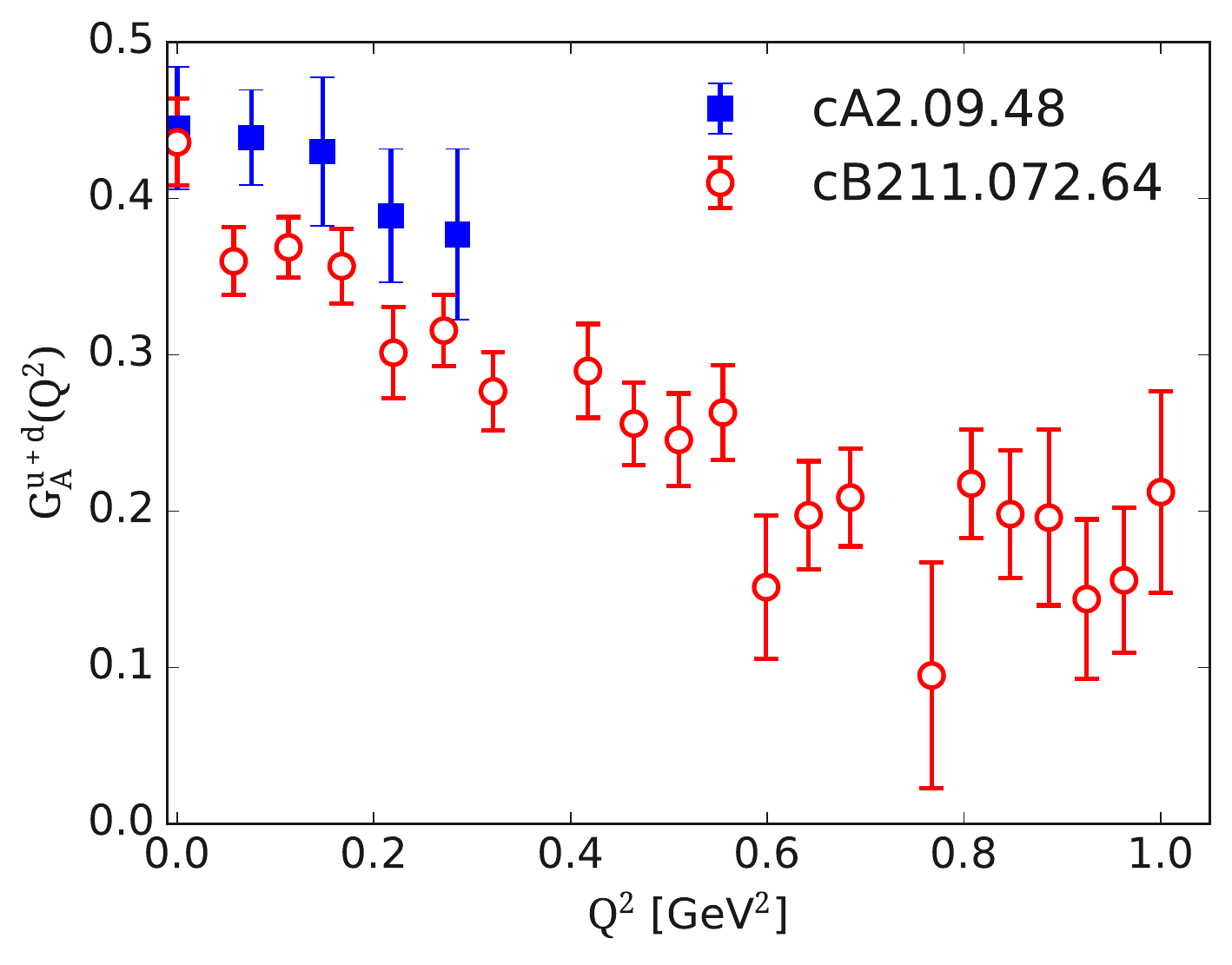}\hspace*{-0.2cm}
     \includegraphics[scale=0.4]{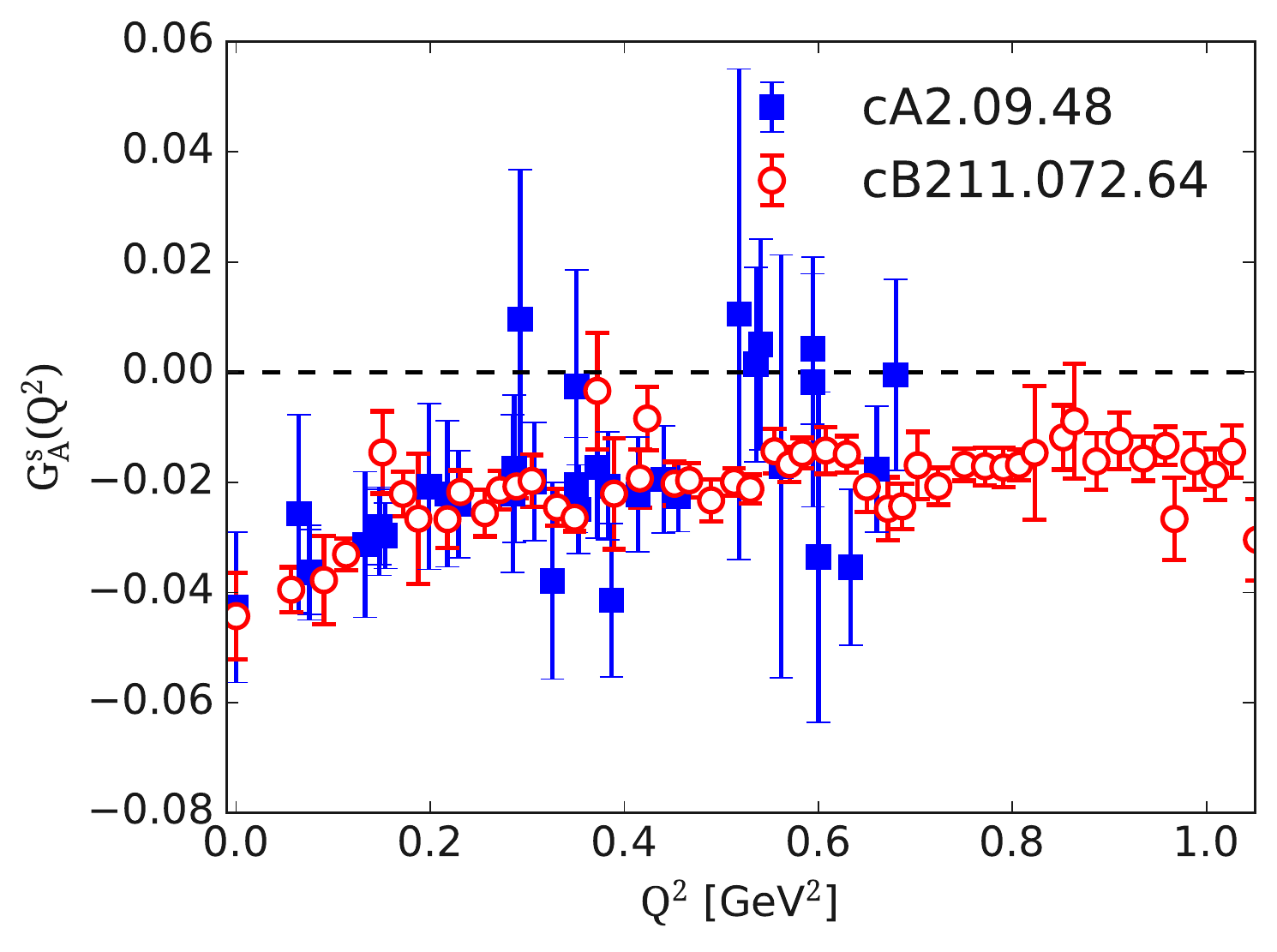}\hspace*{-0.2cm}
     \includegraphics[scale=0.4]{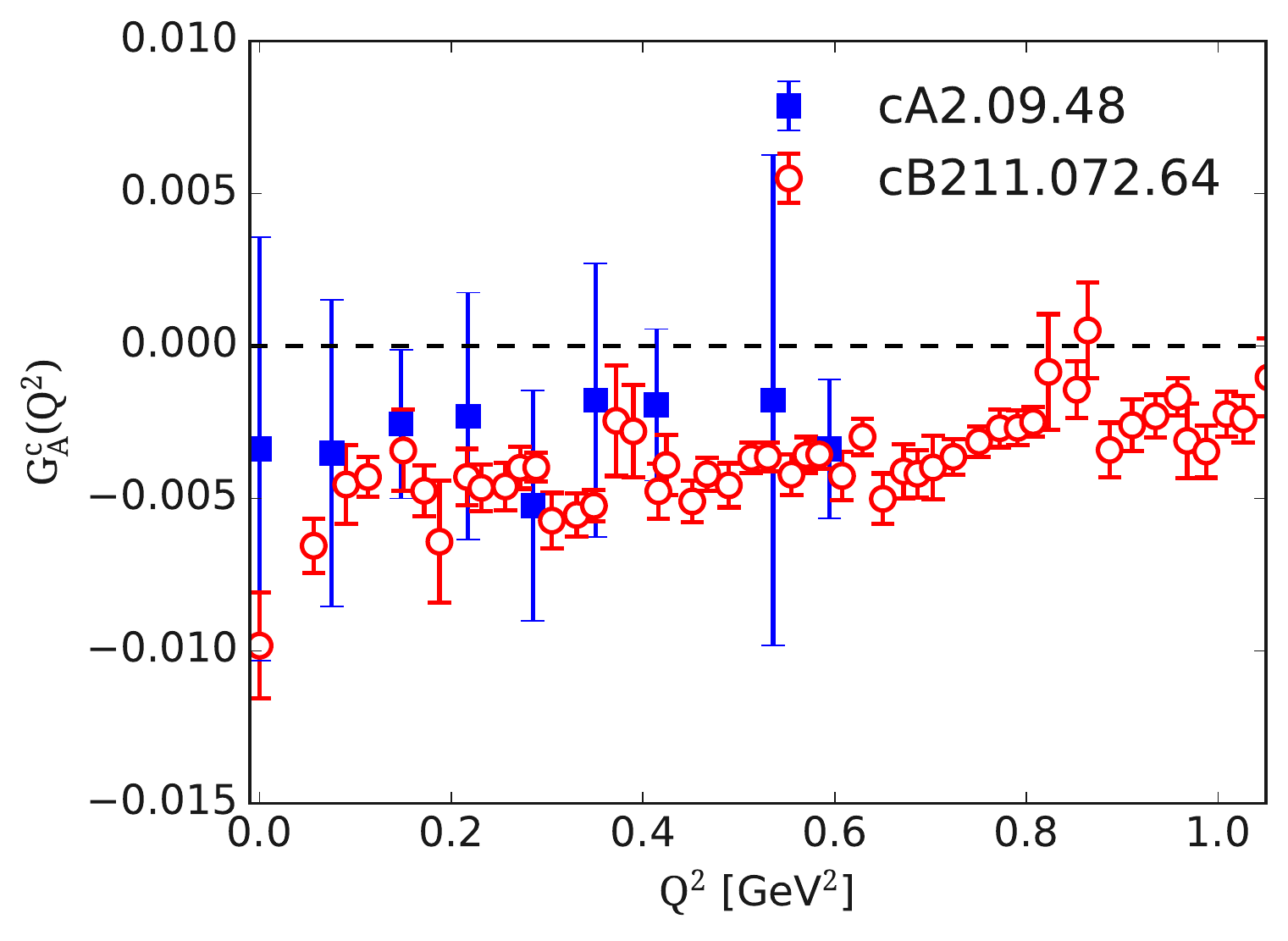}
     \caption{Comparison of the axial form factors using the cB211.072.64 ensemble of this work (open red circles), and the $N_f=2$ ensemble cA2.09.48 (filled blue squares) presented in Ref.~\cite{Alexandrou:2017hac}. See Table~\ref{table:sim} for details on the parameters of the two ensembles. The upper plots show results for  the isoscalar and strange axial form factor while the lower plot for the charm axial form factor. }
     \label{fig:GA_comb}
 \end{figure}
\end{widetext}
 
The form factors presented in this work were studied previously
by only another lattice QCD group, namely the LHPC~\cite{Green:2017keo} but using  an
ensemble with pion mass $m_\pi = 317$~MeV. Here we restrict the comparison to studies performed directly at the physical point and, therefore the only other
available results are provided from our previous work~\cite{Alexandrou:2017hac} 
using the  $N_f=2$ ensemble cA2.09.48 of Table~\ref{table:sim}. In that study we didn't employ the improved noise reduction approaches for the evaluation of the quark loops that we use in this work and presented in Sec.~\ref{ssec:ConnDiscStats}. In our previous study we used volume sources without spin nor color dilution. For the light quarks we used 2250 stochastic sources while for   the strange we used 1024 and for the charm 1250. For the strange and charm quark loops we also used the truncated solver method~\cite{Bali:2009hu}. We note that hierarchical probing used in this work was not used for the analysis of the cA2.09.48.

\begin{widetext}
   \begin{figure}[h!]
     \centering
     \includegraphics[scale=0.4]{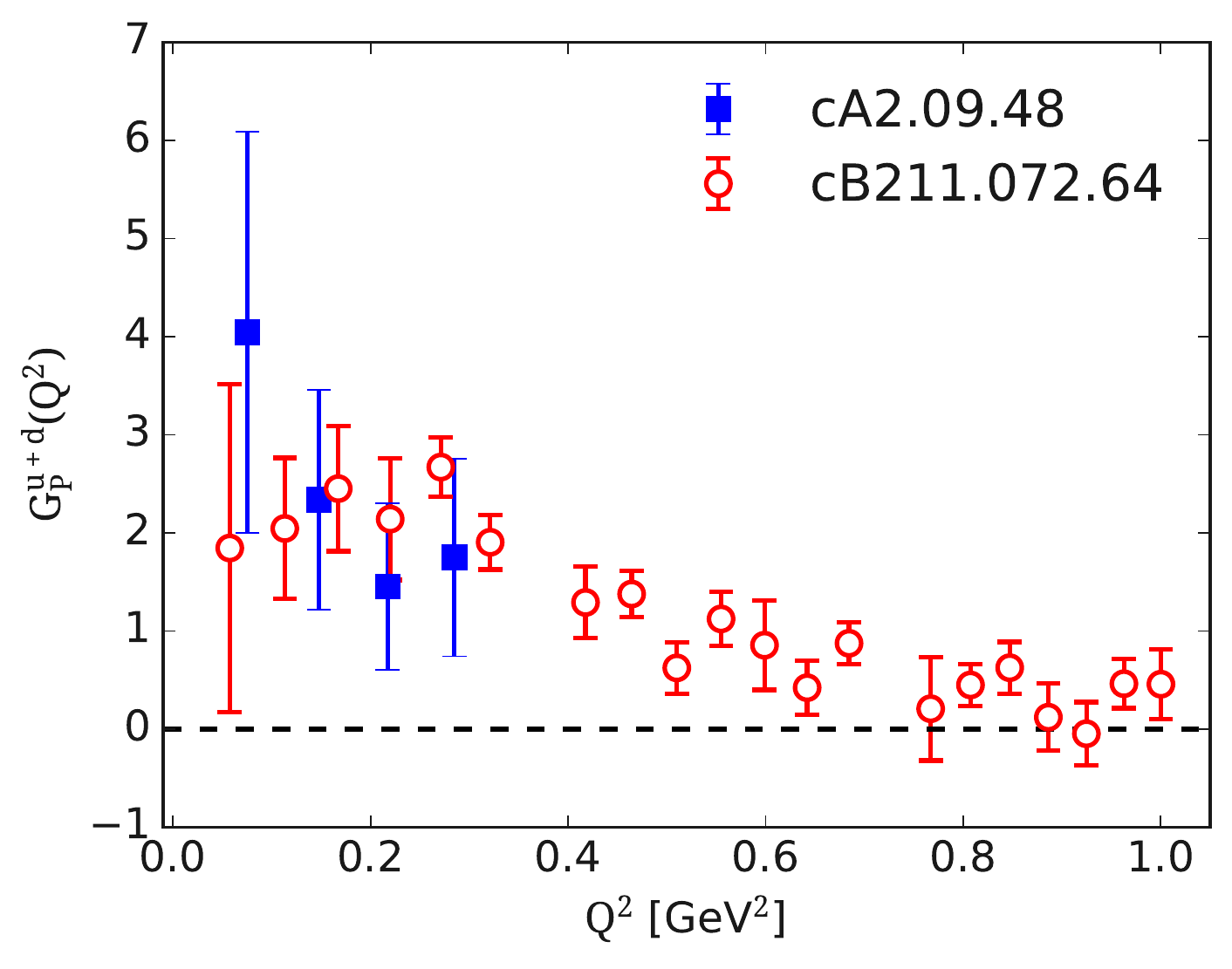}\hspace*{-0.2cm}
     \includegraphics[scale=0.4]{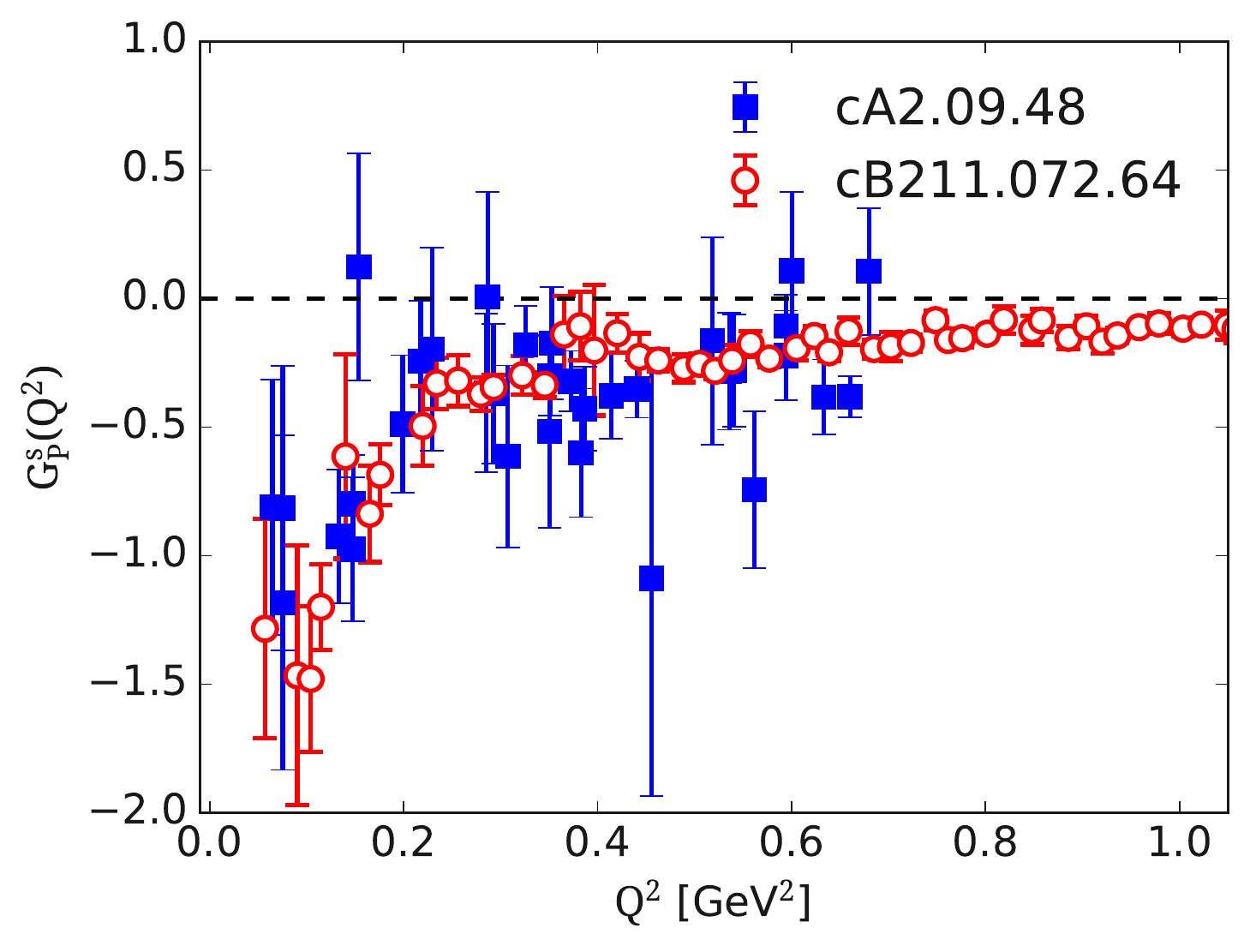}\hspace*{-0.2cm}
     \includegraphics[scale=0.4]{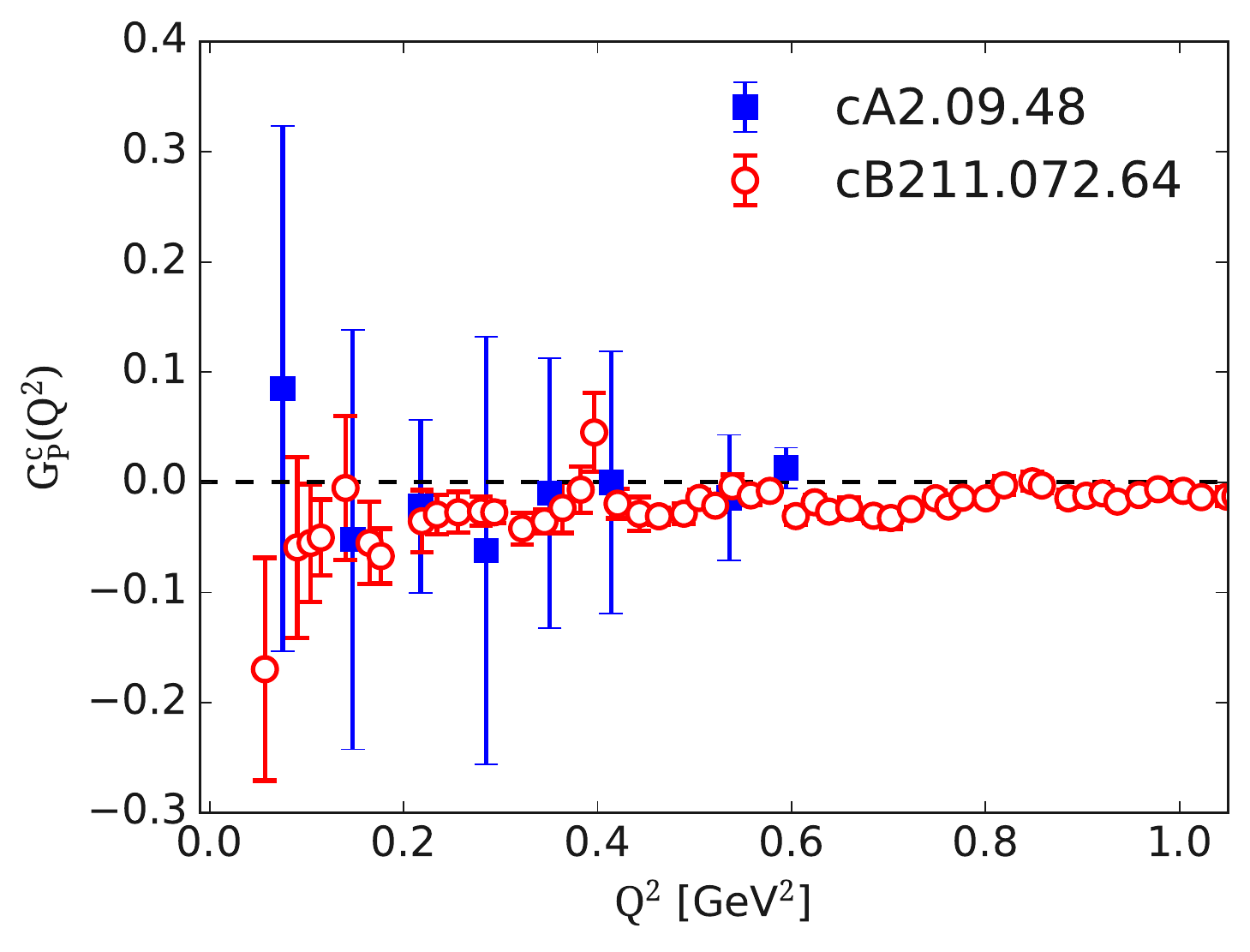}
     \caption{Comparison of the induced pseudoscalar form factors. The notation is as in Fig.~\ref{fig:GA_comb}. 
     }
     \label{fig:GP_comb}
 \end{figure}
\end{widetext} 
 Results for the axial form factors are compared in Fig.~\ref{fig:GA_comb}. For the isoscalar combination only few $Q^2$  values are available in the case of the cA2.09.48 ensemble, namely up to $Q^2=0.3$~GeV$^2$.  For the strange axial form factor, the results using the cA2.09.48 ensemble are very noisy. This comparison provides a nice demonstration of the improvements accomplished in this work with about only twice the computational effort.
 The situation is 
similar for the case of the charm axial form factor.

In Fig.~\ref{fig:GP_comb} we compare the results  for the induced pseudoscalar form factor. We observe agreement between the results using the two ensembles with the results of the current work being significantly more precise.

\section{Conclusions}\label{sec:Summary}
The complete flavor decomposition of the axial and induced pseudoscalar form factors of the nucleon is determined directly at the physical point using one $N_f=2+1+1$ ensemble of twisted mass fermions. We obtain non-zero results for the up, down, strange and charm quark form factors to increased accuracy as compared to our previous study using an $N_f=2$ twisted mass ensemble~\cite{Alexandrou:2017hac}. This is accomplished by using a combination of deflation of lower mode, hierarchical probing, spin-colour dilution and the one-end trick.

These results provide valuable input to on-going and planned parity-violating experiments. They are also crucial for the cross sections for a class of popular cold dark matter candidates ~\cite{Papavassiliou:2009zz}. Having the complete flavor decomposition allows us to check for SU(3) flavor symmetry. We find that SU(3) symmetry is broken up to about 10\% for the octet axial and up to 50\% for the induced pseudoscalar form factors with the breaking being larger at low $Q^2$ values. This is an important result since many phenomenological studies assume SU(3) flavor symmetry, and thus carry an uncontrolled systematic error.

In the future we plan to analyze two additional $N_f=2+1+1$ twisted mass fermion  ensembles with smaller lattice spacings  so that we can take the continuum limit. This will also enable us to check the PCAC relation  directly  in the continuum limit eliminating  any  cut-off effects that may cause violations. 

\begin{acknowledgements}
We would like to thank all members of ETMC for a very constructive and enjoyable collaboration.
M.C. acknowledges financial support by the U.S. Department of Energy, Office of Nuclear Physics, Early Career Award under Grant No.\ DE-SC0020405. 
K.H. is financially supported by the Cyprus Research and Innovation foundation under contract number POST-DOC/0718/0100.
This project has received funding from the Horizon 2020 research and innovation program
of the European Commission under the Marie Sk\l{}odowska-Curie grant agreement No 642069(HPC-LEAP)  and  under grant agreement No 765048 (STIMULATE) as well as  by the DFG as a project under the Sino-German CRC110.
S.B. and J. F. are supported by the H2020 project  PRACE 6-IP (grant agreement No 82376)  and the  COMPLEMENTARY/0916/0015 project funded by the Cyprus Research Promotion Foundation.
The authors gratefully acknowledge the Gauss Centre for Supercomputing e.V. (www.gauss-centre.eu)
for funding the project pr74yo by providing computing time on the GCS Supercomputer SuperMUC
at Leibniz Supercomputing Centre (www.lrz.de).
Results were obtained using Piz Daint at Centro Svizzero di Calcolo Scientifico (CSCS),
via the project with id s702.
We thank the staff of CSCS for access to the computational resources and for their constant support.
This work also used computational resources from Extreme Science and Engineering Discovery Environment (XSEDE),
which is supported by National Science Foundation grant number TG-PHY170022.
We acknowledge Temple University for providing computational resources, supported in part
by the National Science Foundation (Grant Nr. 1625061) and by the US
Army Research Laboratory (contract Nr. W911NF-16-2-0189).
This work used computational resources from the John von Neumann-Institute for Computing on the Jureca system~\cite{jureca} at the research center
in J\"ulich, under the project with id ECY00 and HCH02.
\end{acknowledgements}

\bibliography{refs}

\end{document}